\documentclass[a4paper,11pt]{article}

\usepackage{color}
 \usepackage{a4wide}
\usepackage{siunitx}
\usepackage{stmaryrd}
\usepackage{algorithm,algcompatible}
\algnewcommand\INPUT{\item[\textbf{Input:}]}
\algnewcommand\OUTPUT{\item[\textbf{Output:}]}
\usepackage{graphicx}
\usepackage{subfigure}
\usepackage{multirow}
\usepackage{booktabs}
\usepackage{threeparttable}
\usepackage{amsmath}
\usepackage{amsfonts}
\usepackage{amssymb}
\usepackage{amsthm}
\usepackage{authblk}
\usepackage[round,sort]{natbib}
\usepackage{bm}
\usepackage{caption}
\usepackage{xr}
\usepackage[colorlinks,citecolor=blue,urlcolor=blue]{hyperref}
\usepackage{setspace}

\newcommand{\bfm}[1]{\ensuremath{\mathbf{#1}}}
     \def\bA{\bfm A}     
     \def\bB{\bfm B}

     \def\bQ{\bfm Q}

     \def\bT{\bfm T}     
\def\bu{\bm u}

\def\bx{\boldsymbol x}     \def\bX {\boldsymbol{X}}   
\def\by{\bm y}          
\def\bz{\bm z}     \def\bZ{\bfm Z}     
\def\mP{\mathbb P}

\newcommand{\bfsym}[1]{\ensuremath{\boldsymbol{#1}}}

\def\wh    {\widehat}
\def\wt   {\widetilde}

\def\bbeta     {\bfsym \beta}
\def\btheta     {\bfsym \theta}

\def\bSigma  {\bfsym \Sigma}

\def\bsT{\boldsymbol{\mathcal{T}}}
\def\bsD{\boldsymbol{\mathcal{D}}}

\newtheorem{cond}{Condition}

\makeatletter
\newcommand*{\addFileDependency}[1]{
	\typeout{(#1)}
	\@addtofilelist{#1}
	\IfFileExists{#1}{}{\typeout{No file #1.}}
}
\makeatother

\newcommand*{\myexternaldocument}[1]{%
	\externaldocument{#1}%
	\addFileDependency{#1.tex}%
	\addFileDependency{#1.aux}%
}

\newtheorem{theorem}{Theorem}

\newtheorem{corollary}[theorem]{Corollary}

\providecommand{\keywords}[1]{\textbf{\textit{Keywords:}} #1}

\myexternaldocument{FILTER-supp-V2022}

\title{Modeling High-Dimensional Data with Unknown Cut Points: A Fusion Penalized Logistic Threshold Regression}

\author[1]{Yinan Lin}
\author[2]{Wen Zhou}
\author[3]{Zhi Geng}
\author[4]{Gexin Xiao}
\author[5]{Jianxin Yin \thanks{Corresponding author, jyin@ruc.edu.cn}}

\affil[1]{Department of Statistics and Data Science, National University of Singapore, Singapore, 117546, Singapore}
\affil[2]{Department of Statistics, Colorado State University, Fort Collins, CO, 80523, U.S.A.}
\affil[3]{School of Mathematics and Statistics, Beijing Technology and Business University, Beijing, 100048, China}
\affil[4]{National Institute of Hospital Administration, Beijing, 100044, China}
\affil[5]{Center for Applied Statistics and School of Statistics, Renmin University of China, Beijing, 100872, China}

\begin{document}

\maketitle

\begin{abstract}
In traditional logistic regression models, the link function is often assumed to be linear and continuous in predictors. Here, we consider a threshold model that all continuous features are discretized into ordinal levels, which further determine the binary responses. Both the threshold points and regression coefficients are unknown and to be estimated. For high dimensional data, we propose a  {\bf F}us{\bf I}on penalized {\bf L}ogistic {\bf T}hr{\bf E}shold  {\bf R}egression(FILTER) model, where a fused lasso penalty is employed to control the total variation and shrink the coefficients to zero as a method of variable selection. Under mild conditions on the estimate of unknown threshold points, we establish the non-asymptotic error bound for coefficient estimation and the model selection consistency. With a careful characterization of the error propagation, we have also shown that the  tree-based method, such as CART,   fulfil the threshold estimation conditions. We find the FILTER model is well suited in the problem of early detection and prediction for chronic disease like diabetes, using physical examination data. The finite sample behaviour of our proposed method are also explored and compared with extensive Monte Carlo studies, which  supports our theoretical discoveries.



\end{abstract}

\onehalfspacing

\section{Introduction}\label{s:intro}

\noindent
Discretization is an essential preprocessing technique used in many knowledge discovery and data mining tasks \citep{Alaya:etal:2019, Flores2019, Garcia:etal:2013}. 
Many symbolic data mining algorithms are designed to process such type of discrete data \citep{Garcia:etal:2013, Vollmer:etal:2019}. In \cite{Dussaut:etal:2017}, a binary level discretization using median is naturally appeared, representing the 'activation' and 'inhibition' of genes. 
By using method to split and merge the micro-array continuous concentration data as the discretization method, it is concluded that learning with discrete domains often performs better than the case of continuous data \citep{Sriwanna:etal:2019}. However, several well known simple discretization methods like equal frequency binning (EFB), equal width binning (EWB) and minimum description length principle \citep{Tsai:Chen:2019} does not consider the relevance and mutual information with a supervised response variable.
In \cite{Ferreira:Figueiredo:2015}, features are discretized with relevance and mutual information criteria with respect to a response. 
Also in some cases, features have noisy values or show minor fluctuations that are irrelevant or even harmful for the learning task at hand. For such features, the performance of machine learning and data mining algorithms can be improved by discretization \citep{Franc:etal:2017, Fu:etal:2017}. In \cite{Sokolovska:etal:2018}, a provable algorithm is considered for learning scoring systems with continuous feature binning. In order to reduce time series dimensionality and cardinality, a multi-breakpoints approach is employed to discretize continuous data \citep{Marquez:etal:2020} or some statistical test is employed \citep{Abachi:etal:2018}. Also discrete features are closer to a knowledge-level representation that is easy to understand, use and explain than continuous ones \citep{Tsai:Chen:2019}.

In this paper, the continuous features are discretized, where the cut points(also called threshold points) are supervisedly-learned from data. The discretized features are then plugged into a logistic linear regression model with high dimensional covariates. And a fusion penalty is applied to encourage structured sparse models. Different from the traditional generalized linear model \citep{McCullagh:Nelder:1989}, we allow the discontinuity and non-linear relations between the features and link function. To our best knowledge, the threshold regression model is first considered by \cite{Dagenais:1969} in a setup of time series and the response variable is split into two levels. Specifically, we consider the following model
\[
\mathrm{logit}(p_y) = \beta_0^* + \sum_{j=1}^p \sum_{k=0}^{K_j} \beta_{k,j}^* \mathbb{I}_{\{t_{k,j}^*\le X_j < t_{k+1,j}^*\}},
\]
with a fusion penalty $\mathrm{pen}(\beta, \lambda_n)= \lambda_n\sum_{j=1}^p\sum_{k=1}^{K_j} |\beta_{k,j}-\beta_{k-1,j}|$.  This model can be naturally linked to a risk score derived from the medical examination data. For example, as a chronic metabolic disease which will lead to long term serious damages to many organs, diabetes is becoming worldwide health threat during last decades. Early prediction by scale is very important. Risk scores play a critical role for the early screening and prognosis as well as the prevention and effective treatments of diabetes \citep{Nobel2011}.

We name the above framework the {\bf F}us{\bf I}on penalized {\bf L}ogistic {\bf T}hr{\bf E}shold {\bf R}egression (FILTER) model, whose estimation has been employed to derive a risk score for diabetes based on the physical examination data; see Section \ref{real}. For  ease of presentation, we also call the estimator for our model FILTER. Specifically, the FILTER model splits the samples into multiple regimes according to unknown thresholds of risk factors and regresses the  binary responses  on  indicator functions of these regimes. 

Our contributions about the FILTER model in this paper are two-fold. First, facing two sets of parameters, given the nonparametric rate for the threshold points estimate, FILTER controls the excess risk for predictions base on logistic threshold  regression as well as offers   satisfactory estimations of the regression coefficients. By using the fused Lasso penalty \citep{Tibshirani:etal:2005, Petersen:etal:2016, Tang:Song:2016, wang2018}, FILTER can reduce the total variation for each group of discretized new variables' coefficients.  
Second, we identify a satisfactory  estimator of the unknown thresholds  by CART with the desired non-asymptotic rates. 
Thanks to the discontinuity of the model at threshold points, asymptotic $n$-consistency of estimation was established for both the conditional least squares estimator (CLSE) and maximum likelihood method \citep{Chan:1993, Qian:1998}. Assuming diminishing threshold effect, \cite{Hansen:2000} revealed  that the threshold model reduces to its linear counterpart at rate $n^{-\alpha}$ for $0<\alpha<{1}/{2}$, which yields a slower rate of convergence at $n^{2\alpha-1}$. \cite{Gao:etal:2013} suggested that the $n$-consistency  may not hold with finite samples and the convergence rate should be $T^{-1}(n)$, where $T(n)$ is the number of regenerations in   time interval $[0, n]$ for the $\beta$-null recurrent Markov chains, and $T^{-1}(n)$ is close to $n^{-{1}/{2}}$.

The paper is organized as following. In Section \ref{method.sec}, we introduce the FILTER model, 
and statistical guarantees are obtained for 
the regression coefficients selection and estimation, along with the prediction. A valid threshold point estimation method is analyzed in Section \ref{sec22}.
In Section \ref{simulation.sec}, comprehensive simulation studies are reported to demonstrate the  performance of the proposed method in comparison to competing ones in literature.
Application of the FILTER model to a real physical examination data for diabetes prediction is analyzed in Section \ref{real}. Concluding remarks are given in Sections \ref{last}. More technical results and further simulation studies are deferred to the supplementary materials.

\section{Methodology and Theoretical Properties}
\label{method.sec}

\subsection{The logistic threshold regression model}

Let $\{\bX^{(i)}, y^{(i)}\}_{i=1}^n$  denote independent and identically distributed ({\it i.i.d.}) random samples of $(\bX, y)$, with continuous covariates $\bX \in \mathbb{R}^p$ and label ${y} \in \{-1,1\}$. An alternative yet equivalent label $\{0,1\}$ can be transformed by function $f(y)=(y+1)/2$.
For $p_y=\mP(y=1|\bX)$, termed the probability of success in this paper, consider the model
$
\mathrm{logit}(p_y) = \beta_0^* + \sum_{j=1}^p \sum_{k=0}^{K_j}
\beta_{k,j}^* \mathbb{I}_{\{t_{k,j}^*\le X_j < t_{k+1,j}^*\}},
$
where $\mathbb{I}_{\{\cdot\}}$ is the indicator function,  $\mathrm{logit}(\cdot)$ is the logit function and $t_{k, j}^*$'s are fixed threshold points for $k=0$, $\ldots$, $K_j+1$, $j=1, \ldots, p$. Denoting $\mathcal{X}_j$ as the range of $X_j$, we set $t_{0, j}^*=\inf\{x: x\in \mathcal{X}_j\}$ and $t_{K_j+1,j}^*=\sup\{x: x\in \mathcal{X}_j\}$ for each $j$. In brief, each covariate $X_j$ admits $K_j$ threshold points and $K_j+1$ levels for explaining the variability dwelling in $y$. Letting $Z^*_{k,j} = \mathbb{I}_{\{t_{k,j}^*\le X_j < t_{k+1,j}^*\}}$, the model can be rewritten as
\begin{equation}
	\text{logit}(p_y) = \beta_0^* + \sum_{j=1}^p \sum_{k=0}^{K_j}
	\beta_{k,j}^* Z^*_{k,j}\label{eq01}.
\end{equation}
Conventionally, we assume $\beta_{0,j}^*=0$ for each $j$ to guarantee the identifiablility of the model.
Hence, \eqref{eq01} is the logistic threshold multiple regression model, where the nontrivial nonlinearities implicitly reside in $Z^*_{k,j}$'s. Letting $\bbeta_j^* = (\beta_{1,j}^*, \ldots, \beta_{K_j,j}^*)^{\top}$ and $\bZ^*_j = (Z^*_{1,j}, \ldots, Z^*_{K_j,j})^{\top}$, \eqref{eq01} can be written as
$\text{logit}(p_y) = \beta_0^* + \sum_{j=1}^p \bZ_{j}^{*\top}
\bbeta_{j}^*$. In addition, we assume that there are finite levels for each covariate. That is, $\max_{1\leq j \leq p}K_j \le K_0 < \infty$ for some constant $K_0>0$.
In variable selection regime, it is assumed that there exists a subset $S\subset \{1,\ldots, p\}$ such that $p_y=\mP(y=1|\bX)=\mP(y=1|\bX_S)$, where $\bX_S$ represents the covariates indexed by $S$. We call variables in $S$ the associated variables, and those not in $S$ non-associated variables. Denote the size of $S$ as $|S|=p_0$, where $0< p_0 \le p$. And without loss of generality, assume the first $p_0$ covariates are associated variables, namely $S=\{1,\ldots, p_0\}$, and let $S^{c}$ be its complement. Note that, for $j=1,\ldots, p$, the associated variables are those with non-zero coefficient-vectors $\bbeta_j^*\ne \textbf{0} \in \mathbb{R}^{K_j+1}$, while non-associated variables are those with zero coefficient-vectors $\bbeta_j^* = \textbf{0} \in \mathbb{R}^{K_j+1}$.

Since non-associated variables are with zero coefficient-vectors, \eqref{eq01} is further equivalent to
\begin{equation}\label{eq1}
	\text{logit}(p_y) = \beta_0^* + \sum_{j=1}^{p_0} \sum_{k=0}^{K_j}
	\beta_{k,j}^* Z^*_{k,j}.
\end{equation}
In the above model, both the regression coefficients $\beta_{k, j}^*$ and threshold points $t_{k, j}^*$ in $Z^*_{k, j}=\mathbb{I}_{\{t^*_{k, j}\le X_j<t_{k+1, j}^*\}}$ are parameters to be estimated. But those $t_{k, j}^*$ with $\bbeta_j^*=0$ are not well defined from the estimation perspective. Although there exists methods and algorithms for threshold point $t^*_{k, j}$ estimation from $X_j$ for each $j=1,\ldots, p$, they are not supervised by $Y$ \citep{Garcia:etal:2013}.  
For example, CART (classification and regression tree, \cite{Breiman:etal:1984}) will always give estimates for those $t^*_{k, j}$ of non-associated variables.
We therefore assume that we have estimated threshold points for both associated and non-associated variables. From the results below, we can see that, given the estimated threshold points, FILTER can ensure the coefficients of associated variables to have certain non-asymptotic rate of estimation, as well as consistently estimate the coefficients of non-associated to be zero.

\subsection{The FILTER model and its properties}
\label{sec23}

Denote the response vector by $\by=({y}^{(1)},\ldots,y^{(n)})^{\top}\in \{-1,1\}^n$ and a given generic thresholded design matrix by $\bZ=(\bZ_1, \ldots, \bZ_p)\in \mathbb{R}^{n\times \sum_{j=1}^p
	K_j}$, where $\bZ_j = (Z_{1,j}^{(1)}, \ldots, Z_{K_j, j}^{(1)};
\ldots;$ $Z_{1,j}^{(n)}, \ldots, Z_{K_j, j}^{(n)})\in \mathbb{R}^{n\times K_j}$.  Set $K = \sum_{j=1}^p K_j$, and assume $K_j\ge 1$ for all $j$. Given estimators $\{\wh t_{k,j}\}_{k,j}$ with $\wh t_{0, j}=\inf\{x: x\in \mathcal{X}_j\}$ and $\wh t_{K_j+1,j}=\sup\{x: x\in \mathcal{X}_j\}$ for each $j$, we have
$\wh \bZ_j =(\wh{Z}_{1,j}^{(1)}, \ldots, \wh{Z}_{K_j, j}^{(1)};
\ldots; \wh{Z}_{1,j}^{(n)}, \ldots, \wh{Z}_{K_j, j}^{(n)})\in \mathbb{R}^{n\times K_j}$ and $\wh{\bZ}=(\wh{\bZ}_1,
\ldots, \wh{\bZ}_p)$ with $\wh{Z}_{k,j}^{(i)} = \mathbb{I}_{\{\wh t_{k,j}\le X_j^{(i)} < \wh t_{k+1,j}\}}$. For $j \in S$, $K_j$, given before estimation, is the number of threshold points.
With a generic thresholded design matrix $\bZ$, to encourage the continuity of  risk score with respect to the adjacent levels of risk factors and reduce the total variance of the coefficients, we consider the objective function with the fusion penalty
\begin{equation}
	\mathcal{S}(\bB; \by, \bZ) = \mathcal{L}(\bB;
	{\by}, \bZ) + \lambda_n
	\sum_{j=1}^p\sum_{k=1}^{K_j}
	|\beta_{k,j}-\beta_{k-1,j}|\label{objfunc:fusedlasso},
\end{equation}
where $\bB=(\bbeta_1^{\top},\ldots, \bbeta_p^{\top})^{\top}$ with $\bbeta_j=(\beta_{1,j},\ldots, \beta_{K_j,j})^{\top}$, $\beta_{0,j}=0$ for each $j$, and $K_j$ may vary across $j$'s. In addition, the negative log-likelihood for the logistic regression is
\begin{align}
	\notag
	\mathcal{L}(\bB;{\by}, \bZ) &=
	-\sum_{j=1}^p\langle \bbeta_j, \frac{1}{n}\sum_{i=1}^n y^{(i)}
	\textbf{Z}_j^{(i)} \rangle + \frac{1}{n}\sum_{i=1}^n
	\psi\left(\sum_{j=1}^p\langle \bbeta_j, \bZ_j^{(i)} \rangle\right) \\
	&= - \frac{1}{n}\sum_{i=1}^n {y}^{(i)} \langle \bB,
	\textbf{Z}^{(i)} \rangle + \frac{1}{n}\sum_{i=1}^n
	\psi(\langle \bB, \textbf{Z}^{(i)} \rangle),
	\label{logisticloss}
\end{align}
with $\psi(u) = \log(\exp(u)+\exp(-u))$, and $\bZ^{(i)}$ the $i$th row of $\bZ$. It can be seen that the shape of $\bB$ is associated with the shape of $\bZ$. Consequently, if we solving \eqref{objfunc:fusedlasso} with the true thresholded design matrix $\bZ^*$, the estimated coefficients vector may have a different shape with the one from the estimated thresholded design matrix $\wh \bZ$.


To establish the statistical guarantees on $\wh \bB$, for a given thresholded design matrix by $\bZ$, denote $\bsT =
\text{diag}(\bT_1,\ldots, \bT_p)$ the diagonal block matrix with $K_j\times K_j$ diagonal matrices $\bT_j$'s, where the difference matrix $\bT_j=(\mathsf{t}_{j,k\ell})_{1\leq k,\ell\leq K_j}$ has $\mathsf{t}_{j,k\ell}=1$ if $k=\ell$, $\mathsf{t}_{j,k\ell}=-1$ if $k-\ell=1$, otherwise $\mathsf{t}_{j,k\ell}=0$. Note here $\bsT$ may have different shapes with different $\bZ$'s. The shape of $\bsT$ can be deduced from the context. It can be seen that $\bsT$ is invertible and $\bsD=\bsT^{-1}=\text{diag}(\bT_1^{-1},\ldots, \bT_p^{-1})$. Letting $\btheta=\bsT\bB$, $\wt{\bZ}=\bZ\bsD$ that $\wt{\bZ}^{(i)}=\bsD^{\top}\bZ^{(i)}$, and $\wt{\by}=\by$, \eqref{logisticloss} is then rewritten as
$
\mathcal{L}(\btheta;\wt{\by}, \wt{\bZ})  =
- n^{-1}\sum_{i=1}^n {y}^{(i)} \langle \btheta,
\wt{\bZ}^{(i)} \rangle + n^{-1}\sum_{i=1}^n
\psi(\langle \btheta, \wt{\bZ}^{(i)} \rangle),
$
so that \eqref{objfunc:fusedlasso} with penalty  $\lambda_n \|\bsT \bB\|_1$ is equivalent to
\begin{equation}
	\mathcal{S}(\btheta;\wt{\by}, \wt{\bZ}) = \mathcal{L}(\btheta;\wt{\by}, \wt{\bZ}) + \lambda_n \|\btheta\|_1. \label{transloss}
\end{equation}
That is, the FILTER model is reduced to the $\ell_1$-regularized logistic regression, which has been widely studied \citep{Ravikumar:etal:2010, Buhlmann:Sara:2011}. In addition, the FILTER model is also similar to the predictor-corrector method introduced by \cite{Park:Hastie:2007} to learning the Lasso path for generalized linear models.



By model \eqref{eq1}, we have the corresponding population version of true thresholded covariates of associated variables $\bz^*_{S}=(\bz^*_1, \ldots, \bz^*_{p_0})\in \mathbb{R}^{n\times (\sum_{j=1}^{p_0}
	K_j)}$ with $\bz^*_j = (z^*_{1,j}, \ldots, z^*_{K_j, j})^\top\in \mathbb{R}^{K_j}$ for $j\in S$.
Let $\bB^*_{S}=(\bbeta_1^{*\top},\ldots, \bbeta_{p_0}^{*\top})^{\top}$ be the true regression coefficients of associated variables corresponding to $\bz^*_S$, where $\bbeta_j^* = (\beta_{1,j}^*, \ldots, \beta_{K_j,j}^*)^{\top}$, and $y$ be the population of response variable. Moreover, let $\mathsf{z}^{*}_{S}$ be the population of the true thresholded covariates with levels of associated variables centered, namely
\[
\mathsf{z}^{*}_j = \bz^{*}_j - \mathbb{E}[\bz^{*}_j], \qquad j\in S,
\]
and $l(\bB^*_{S};y, \mathsf{z}^{*}_{S}) = -y\bB^*_{S}\mathsf{z}^{*}_{S} + \psi(\bB^*_{S}\mathsf{z}^{*}_{S})$, then the Fisher information matrix of centered thresholded associated covariates is
$\bQ^{*}_{SS}= \mathbb{E}[\nabla^2 l(\bB^*_{S};y, \mathsf{z}^{*}_{S})]= \mathbb{E}
[\eta(\mathsf{z}^{*}_{S};\bB^*_{S})\mathsf{z}^{*}_{S} \mathsf{z}_{S}^{*\top}]$, where $
\eta(\bu;\bB) = {4\exp(2\bB^{\top} \bu)}\{\exp(2\bB^{\top} \bu) + 1\}^{-2}$. Denote $S_{\bB^{*}}$ the support of
$\bB^{*}_{S}$ with size $|S_{\bB^{*}}|=d/2$ and $S_{\bB^{*}}^c$ its complement. Note here $\bB^{*}_{S}$ may be different with $\bB^{*}_{S_{\bB^{*}}}$, since there may be some levels of some associated variables being zeros. Let $\bQ^{*}_{S_{\bB^{*}}S_{\bB^{*}}}$ be the $d/2\times d/2$ sub-matrix of $\bQ^{*}_{SS}$ whose indices of rows and columns belong to $S_{\bB^{*}}$. We impose some regularity conditions below to study the statistical properties of FILTER. In this paper, we assume all covariates $X_1$, $\ldots$, $X_p$ are continuous.

\begin{cond}[Dependency] There exist constants $C_{\min}>0$ and $ D_{\max}<\infty$ such that $\Lambda_{\min}(\bQ^{*}_{S_{\bB^{*}}S_{\bB^{*}}}) \ge C_{\min}$ and $\interleave \mathbb{E}[\mathsf{z}^{*}_{S} \mathsf{z}^{*\top}_{S}] \interleave_2 \le D_{\max}$, where $\Lambda_{\min}(\bA)$ and  $\interleave \bA \interleave_{2}$ denote the smallest eigenvalue and  spectral norm of matrix $\bA$, respectively.
	\label{c1}
\end{cond}

\begin{cond}[Incoherence] There exist an $\alpha \in (0,1]$ and an $\alpha_0 \in [1-\frac{1-\alpha}{2K_0}, 1]$ such that
	\[
	\sum_{a=1}^{p_0}\sum_{b=0}^{K_a}       \left|\sum_{j=1}^{p_0}\sum_{l=0}^{K_j}\mathbb{E}[\eta(\mathsf{z}^{*}_{S};\bB^*_{S})\mathsf{z}^{*}_{l,j}]q_{(j-1)K_j+l,(a-1)K_a+b}\right| \le 1 - \alpha_0,
	\]
	where $q_{m,n}$ is the element of $(\bQ^{*}_{S_{\bB^{*}}S_{\bB^{*}}})^{-1}$.
	\label{c2}
\end{cond}

\begin{cond}[Mixing]\label{c13}
	There exists an $0<\epsilon \le 1$, such that
	\[
	\sup_{\substack{m=0,\ldots,K_n\\n\in S^c }} \left| \mathbb{P}\left(X_n\in T_{m,n}|X_j\in T_{l,j}^*,l=0,\ldots,K_j,~j\in S\right) - \mathbb{P}\left(X_n\in T_{m,n}\right) \right| \le \epsilon,
	\]
	where $T_{l,j}^*=(t_{l,j}^{*}, t_{l+1,j}^{*}]$ for $l=0,\ldots,K_j,~j\in S$ by true threshold points, and $T_{m,n}=(t_{m,n}, t_{m+1,n}]$ for $m=0,\ldots,K_n,~n\in S^c$ by any estimated threshold points $t_{m,n}$.
\end{cond}

\begin{cond}\label{c9}
	Let $d = 2|S_{\bB^{*}}|$, $S_{\btheta^{*}}$ be the support of $\btheta^{*}=\bsT\bB^{*}$, $\theta^{*}_{\min} = \min_{j\in S_{\btheta^{*}}} |\theta^{*}_j|$ and $\theta^{*}_{\max} = \max_{j\in S_{\btheta^{*}}} |\theta^{*}_j|$. Assume
	$\theta^{*}_{\max}\le C_0$ for some constant $C_0>0$, and $|\theta^*_{\min}|\ge {10}{C_{\min}}^{-1}\sqrt{d}\lambda_n$.
\end{cond}


\begin{cond}\label{c10}
	Given estimators of threshold points $\{\wh t_{k,j}\}_{k,j}$, for associated variables $j\in S$, with probability at least $1-b_n$, we have $|\wh{t}_{k,j} - t_{k,j}^*| \le a_n$, $k=0,\ldots, K_j$, where $a_n$ and $b_n$ should be positive sequences converging to 0. In addition, $a_n n \to \infty$ and $b_n$ converges exponentially fast as $n\to \infty$.
\end{cond}

\begin{cond}\label{c11}	With $C_1\geq \max\{4 ({\theta}_{\max}^{*}+\frac{C_{\min}}{2D_{\max}}), 1\}$,
	\[
	\lambda_n \ge \frac{(2-\alpha)16}{\alpha} \max\left\{\sqrt{\frac{\log p}{n}}, \frac{5}{2}C_1dK_0 f^{\max}a_n^{1/2}\right\},
	\]
where $d$ is in Condition \ref{c9}, $\alpha$ is in Condition \ref{c2}, $a_n$ is in Condition \ref{c10}, and $f^{\max} = \sup_{j=1,\ldots, p} f_j^{\max}$ with $f_{j}^{\max}$ being the maximum value of the $X_j$'s density function.
\end{cond}

\begin{cond}\label{c14} Let constant $L>0$ be free from $(n, p, d)$, with $d$ in Condition \ref{c9}, $a_n$ and $b_n$ in Condition \ref{c10},  and $\tau_n$ being the smallest $n$ such that $a_n < \frac{1}{2}\min\{t^*_{k+1,j}-t^*_{k,j}, k=0,\ldots,K_j, j\in S\}$, it holds
	\[
	n \lambda_n^2 / d^2 \to \infty,~db_n=o(1),~ \sqrt{d}\lambda_n=o(1),~ n>\max\{L d^3 \log p, \tau_n\}.
	\] 
\end{cond}

Conditions \ref{c1} and \ref{c2} essentially come from \cite{Ravikumar:etal:2010} with the adjustment for centered population.
Moreover, Condition \ref{c2} is the irrepresentable condition \citep{Zhao:Yu:2004} or the incoherence condition \citep{Wainwright:2009}, and either of them assume that the associated variables have small correlation with the non-associated variables. In our setting, due to the absence of the true threshold points for the non-associated variables, the non-associated variables are not well defined and not "observed" in the model. The incoherence condition via the block-wise sub-matrix form containing information from the non-associated variables as in \cite{Ravikumar:etal:2010} cannot be applied directly. Therefore, as a complement to Condition \ref{c2}, we impose Condition \ref{c13}, one kind of $\phi$-mixing condition, to require the weak correlations between associated variables and non-associated variables. Conditions \ref{c2} and \ref{c13} will lead to the mutual incoherence condition in \cite{Ravikumar:etal:2010}.
Besides, Condition \ref{c2} is also adjusted with an extra requirement. This comes from the additional transformation $\mathcal{D}$ with $\interleave \mathcal{D} \interleave_{\infty} \le K_0$ and $\interleave \mathcal{D}^{-1} \interleave_{\infty} \le 2$ in our setting. For more details, see the proofs in the supplementary materials.

Condition \ref{c13} is equivalent to
\[
\sup_{\substack{m=0,\ldots,K_n\\n\in S^c }} \left| \mathbb{E}\left(\wh\bz_{m,n}|\bz_{l,j}^{*},l=0,\ldots,K_j,~j\in S\right) - \mathbb{E}\left(\wh\bz_{m,n}\right) \right| \le \epsilon,
\]
where $\bz_{l,j}^{*}$ is with respect to the $l$th level of the $j$th variable in the uncentered population of the true thresholded covariates $\bz^{*}$ for associated variable $j\in S$, and $\wh\bz_{m,n}=\mathbb{I}_{\{ t_{m,n}\le X_n < t_{m+1,n}\}}$ is the uncentered population of the estimated thresholded covariates with any $t_{m,n}$ for non-associated variable $n\in S^c$. We shall use this form in our proofs, and we provide some examples satisfying Condition \ref{c13} in the supplementary materials.


Condition \ref{c9} specifies the minimal signal level to recover the regression coefficients. Moreover, it requires that the maximal difference of adjacent levels should not be upper bounded, which is true for fixed coefficients. Such a requirement ensures the probability of success not being extreme large or small, meaning the true model is nearly singular, similar conditions are adopted by \cite{Bach:2010} and \cite{bunea2008honest}. A more comprehensive study for the effect of the magnitude of coefficients on the existence of maximum likelihood estimation for the logistic regression can be found in \cite{sur2019modern}. 
We assume all variables are continuous without loss of generality, for discrete nominal and ordinal features are more easy to process.
Condition \ref{c10} imposes the rate the estimated threshold points should have. For tuning parameter $\lambda_n$, Condition \ref{c11} imposes the constraints both from the usual rate $\sqrt{\log p/n}$ to dominant noise in high dimensional predictors and to cover the error rate of threshold points estimation.
Finally, Condition \ref{c14} specifies the configuration to ensure the consistency in the following Theorem \ref{thm4}.

For the traditional linear or logistic regressions,   conditions similar to Conditions \ref{c1} to \ref{c13} are imposed for the control of  correlations among covariates. But in our setting, threshold points also matter. Therefore, Conditions \ref{c1} to \ref{c13}  reflect the sophistication of the FILTER model due to unknown threshold points. 
In addition, our model includes the Ising model as a special case \citep{Ravikumar:etal:2010}. Actually, if there is only one level with non-zero true coefficient for each covariate, i.e., the response and covariates are all binary, our model reduces to the Ising model. Therefore, when the assumptions (A1) and (A2) in \cite{Ravikumar:etal:2010} are satisfied, Conditions \ref{c1} and \ref{c2}  also hold. Moreover, with examples satisfying   Condition \ref{c13}   in the supplementary materials, Conditions \ref{c1} to \ref{c13} are not unrealistic in practice.

We are in position to establish the consistency on both estimating and selecting the regression coefficients for the FILTER model. Recall $S\subset \{1,\ldots, p\}$ is the index set of all associated variables, and $|S|=p_0$. Moreover, let $[S]$ be the column indices of all associated variables' levels in $\bZ^*_{S}$, the design matrix consists of all associated variables, and the size of $[S]$ is $\sum_{j\in S} K_j$.

\begin{theorem}
	For Model   \eqref{eq1} with thresholded design matrix $\bZ^*_{S}$ and regression coefficients $\bB^*_{S}$, assume Conditions \ref{c1} to \ref{c14} hold. Especially, the threshold points $\widehat t_{k, j}$s for each covariates satisfy Condition \ref{c10}. Consider
	\begin{equation}\label{argmin1}
		\wh{\bB} \in \mathop{\mathrm{argmin}}\limits_{\bB\in\mathbb{R}^{K}}
		\mathcal{S}(\bB; \by, \wh{\mathsf{Z}}),
	\end{equation}
	where $\wh{\mathsf{Z}}=\wh{\bZ}-\mathbb{E}_n[\wh{\bZ}]$ is the centered estimated thresholded design, where $\mathbb{E}_n$ denotes for the empirical mean. Then there exists a constant $M>0$ such that the following properties
	hold with probability at least $1-\exp\left\{ -M n \lambda_n^2/p_0^2 \right\} - 2p_0K_0b_n - 2p_0K_0 \exp\{-Mna_n\}$,
	\begin{enumerate}
		\item ({Sign consistency}) Problem in \eqref{argmin1} admits a unique solution and for associated variables in $S$,
		$\wh{\btheta}_{[S]}$ correctly select all the non-zero components
		of $\btheta^*_{[S]}$. Moreover, it has sign-consistency, namely
		$\textrm{sign}(\wh{\btheta}_{[S]})=\textrm{sign}(\btheta^*_{[S]})$. While for each non-associated variable $j \in S^c$, $\wh{\btheta}_{k,j}=0, k=1,\ldots, K_j$. Thus, for $j \in S^c$, no matter what $\{\wh t_{k,j}\}$ are, the corresponding coefficients can always be merged into one 0 over the whole domain of each non-associated variable.
		\item ({$\ell_2$-consistency}) 
		$\|\wh{\bB}_{[S]}- \bB^*_{[S]}\|_2 \le {5K_0}{C_{\min}}^{-1} \sqrt{d}\lambda_n$, where $\|\cdot\|_2$ is the $\ell_2$-norm of a vector.
	\end{enumerate}
	\label{thm4}
\end{theorem}

The sign consistency on the difference of adjacent regression coefficients encourages the desired continuity in the FILTER model. A byproduct of such a continuity is the model selection of $X_j$'s.
As $\beta_{0,j}^*=0$ for each $j$, covariates with all adjacent differences identified as zero, {\it i.e.}, $\theta_{k,j}=0$ for $k=1,\ldots, K_j$, will not be picked.
Also, similar to \cite{Ravikumar:etal:2010}, conditions in Theorem \ref{thm4} are imposed on the design matrix. However, the design matrix for FILTER model is thresholded
and depends on estimated threshold points, which results in a {\it de facto} errors-in-variable setting. This leads to the modified $\lambda_n$ compared to Lasso, as well as the extra term $2p_0K_0b_n$ in the overwhelming probability in the above theorem. For another term $2p_0K_0 \exp\{-Mna_n\}$ in the overwhelming probability, it comes from the centering step for $\wh{\mathsf{Z}}$, as the centering is based on the estimated threshold points instead of true threshold points. 

As we will see in Section \ref{sec22}, a CART-type estimators for the threshold points shall
share $a_n = n^{-1/2+\delta_0}$ and $b_n=2\exp(-n^2/\delta)$ for some $1/2>\delta_0>\delta>0$ in Conditions \ref{c10} to \ref{c14}.
This guarantees the existence of a satisfactory  estimator for the unknown threshold points.

Due to Condition \ref{c14}, the probability in the Theorem \ref{thm4} is tending to one as $n$ goes to infinity, while the upper bound in ($\ell_2$-consistency) is tending to zero. 
When the number of associated variables $p_0$ (or equally, $d=2p_0$), is bounded or grows slowly enough, Condition \ref{c14} allows
the {\it large $p$ and small $n$} setting. 
For example, we consider a setting with
$d = \Theta (n^{c_1})$ and $p = \Theta (e^{n^{c_2}})$, with $0<c_1<1/10$, $2c_1+1/2<c_2<1-3c_1$, 
where $\Theta(\cdot)$ means the same order. 
We are now raising an example that conditions \ref{c11} and \ref{c14} are satisfied. Since there are two parts in $\lambda_n$ in Condition \ref{c14}, to simplify the requirement for $\lambda_n$, we consider the case $a_n = n^{-1/2+\delta_0}$ with $\delta_0=c_2-2c_1-1/2$($>0$). This convergence rate $a_n$ is satisfied by the CART-type estimator, which is proved in Section \ref{sec22}. 
With such an $a_n$, we have $\Theta(d a_n^{1/2})\lesssim \Theta(\sqrt{\log p / n})$, where $\lesssim$ means the smaller or the same order.
In this case, Condition \ref{c11} can be simplified as $\lambda_n \gtrsim \Theta(\sqrt{\log p / n})$.
With this simplification, we now turn to consider Condition \ref{c14}. Firstly, $db_n=o(1)$ is satisfied automatically, as $b_n$ is assumed to converge exponentially fast in $n$. Secondly, it can be seen that 
$n\lambda_n^2 / d^2 \gtrsim n^{c_2-2c_1} \to \infty$, 
while $d^3 \log p =\Theta(n^{3c_1+c_2})=o(n)$ as required in Condition \ref{c14}. 
Moreover, taking $\lambda_n = \Theta(\sqrt{\log p / n})$, we have $\sqrt{d}\lambda_n = o(1)$, the last requirement in Condition \ref{c14}. 
Consequently, 
$
\|\wh{\bB}_{[S]}- \bB^*_{[S]}\|_2 \lesssim \Theta(\sqrt{d\log p/n}) = \Theta(n^{(c_1 + c_2 - 1)/2}) = o(1).
$
Such an upper bound for error is common in the high-dimensional statistics.

In addition, with this high-dimensional setting, the quantity $|\theta^*_{\min}|$,
the smallest signal level in Condition \ref{c9} to recover the signs of the true model, can be lower bounded
\[
|\theta^*_{\min}| \gtrsim \Theta\left(\sqrt{\frac{d\log p}{n}}\right) = \Theta\left(n^{\frac{c_1+c_2-1}{2}}\right).
\]

As a straightforward corollary, Theorem \ref{thm4} also leads to the consistency for predictions in $\ell_2$ sense.

\begin{corollary}
	\label{cy1}
	Let $\beta^*_{\max} = \max_{\{k,j\}} |\beta^*_{k,j}|$. Under conditions in Theorem \ref{thm4}, we have
	\begin{equation}
		\frac{1}{ \sqrt{n}} \| \wh{\bZ}\wh{\bB} -  \bZ^*_{S}\bB^*_{S} \|_2 \le \left(\frac{5}{C_{\min}}\lambda_n + \beta_{\max}^*\right) \frac{K_0 d^{\frac{3}{2}}}{\sqrt{n}},
	\end{equation}
	with overwhelming probability.
\end{corollary}

Corollary \ref{cy1} characterizes the prediction of the FILTER model. If we further require $a_n \gtrsim \Theta(n^{-1/2})$, the error upper bound will tend to zero. Indeed, by the requirements for $\lambda_n$ in Condition \ref{c11}, we always have $\lambda_n\gtrsim \Theta(d a_n)$. In order to satisfy   $\sqrt{d}\lambda_n=o(1)$ in Condition  \ref{c14}, we need $\Theta(d a_n)\lesssim \lambda_n \lesssim \Theta(d^{-1/2})$, resulting $a_n \lesssim \Theta(d^{-3/2})$. In all, $\Theta(n^{-1/2}) \lesssim a_n \lesssim \Theta(d^{-3/2})$ is satisfied once $d \lesssim o(n^{1/3})$, 
giving the convergence rate of the main term in the error upper bound being $d^{3/2}/\sqrt{n}=o(1)$. In addition, with the high-dimensional setting after the Theorem \ref{thm4}, we have
\[
\frac{1}{\sqrt{n}} \| \wh{\bZ}\wh{\bB} -  \bZ^*_{S}\bB^*_{S} \|_2 \lesssim \Theta\left(n^{-\frac{c_2}{2}}\right) = o(1).
\]

As we can see, the error upper bound from the above corollary is mainly due to the second term $\beta_{\max}^* K_0 d^{3/2}n^{-1/2}$, which is unrelated to $\lambda_n$. This is because there are some estimation errors from the estimated threshold points in the estimated design matrix $\wh{\bZ}$ for prediction. Consequently, this makes the prediction errors different from the common convergence rate in the high-dimensional statistics.

Finally, we conclude this section by investigating the   excess risk for prediction of the FILTER model.
For Model \eqref{eq1} with regression coefficients $\bB^*_{S}$ and uncentered thresholded covariates $\bz^*_{S}$ with true threshold points $\{t_{k,j}^*\}$ based on a input $\bx \in \mathbb{R}^p$, denote the corresponding Bayes classifier by $g^*$. Let $\eta^*(\bx) = \{1+\exp(-(\bz^*_{S})^{\top}\bB^*_{S})\}^{-1}$, then,
\begin{equation}\label{bayesp}
g^*(x) = \left\{
\begin{aligned}
	1 &\quad \text{if } \eta^*(\bx)>1/2,\\
	-1 &\quad \text{otherwise}.
\end{aligned}
\right.
\end{equation}\
For new input $\bx \in \mathbb{R}^p$, the prediction by FILTER  $\wh{g}$ replaces $\eta^*$ in \eqref{bayesp} by  $\wh{\eta}(x) =\{1+\exp(-\wh{\bz}^{\top}\wh{\bB})\}^{-1}$, where 
$\wh{\bB}$ and  $\wh{\bz}$ are the  estimated regression coefficients and thresholded covariates derived from $\{\wh{t}_{k,j}\}$. The excess risk of FILTER is 
$
\mathcal{E}(\wh{g}, g^*) = L(\wh{g}) - L(g^*),
$ where $L(g)=\mathbb{P}\{g(\bX)\ne y\}$ is the error probability of prediction $g$.

\begin{cond}\label{c12}
	There exists $C_0 > 0$ such that $|(\bz^*_{S})^{\top}\bB^*_{S}|\le C_0$ for the true thresholded covariate $\bz^*$.
\end{cond}

Condition \ref{c12} assumes that the magnitude of the maximum in each block of $\bB^*_{S}$ is not too large, with which, we have the following.
\begin{theorem}
	Under the conditions in Theorem \ref{thm4}, for a new sample $\bX\in \mathbb{R}^p$ with true thresholded covariate $\bz^*_{S}$ satisfying Condition \ref{c12}, we have
$
	\mathcal{E}(\wh{g}, g^*) \le C_1 d \lambda_n + C_2 d^{\frac{3}{2}} a_n,
$
	for some universal constants $C_1, C_2$.
	\label{thm5}
\end{theorem}

Theorem \ref{thm5} suggests that, whenever $d=\Theta(1)$, the excess risk $\mathcal{E}(\wh{g}, g^*)$   shrinks to zero as $n$ goes to infinity. When $d$ diverges, if we strengthen the requirement $\sqrt{d}\lambda_n=o(1)$ in Condition \ref{c14} to the requirement $d\lambda_n=o(1)$,    the excess risk still shrinks to zero.
With such a modification,   $C_1 d \lambda_n$   will converge to zero. Additionally, as $\lambda_n\gtrsim \Theta(d a_n)$ and $d\lambda_n=o(1)$,
we have $\Theta(d a_n)\lesssim \lambda_n \lesssim \Theta(d^{-1})$, resulting in $a_n \lesssim \Theta(d^{-2})$. Thus, the second term in the upper bound $d^{3/2} a_n \lesssim \Theta(d^{-1/2})=o(1)$. As an example, with the high-dimensional setting after   Theorem \ref{thm4}, we have
$
d \lambda_n \lesssim \Theta(n^{(2c_1+c_2-1)/2}), ~\text{and}~ d^{3/2}a_n \lesssim \Theta(d^{-1/2}) = \Theta(n^{-{c_1}/{2}}).
$
As $3c_1+c_2 < 1$, $\Theta(n^{(2c_1+c_2-1)/2}) \lesssim \Theta(n^{-{c_1}/{2}})$.  Theorem \ref{thm5} therefore bounds the convergence rate of the excess risk as $\Theta(n^{-{c_1}/{2}})=o(1)$. In summary, 
 as a classifier, the FILTER model enjoys the classification consistency, which demonstrates the effectiveness of the FILTER model.

\section{Estimating threshold points using CART}
\label{sec22}
As suggested by Condition \ref{c10}, the estimation of the FITLER model and corresponding predictions depend on the recovery of unknown threshold points. In this section, we identify   a satisfactory statistical procedure for that purpose, which enjoys both desired guarantees and computational efficiency. 
For ease of exposition, we assume $K_j=1$ for each $j$ in this section while the generalization to $K_j\geq 2$ is straightforward.
Thanks to its additive nature, the model \eqref{eq1} is essentially a collection of adjacent hypercubes in $\mathbb{R}^p$ with aligned edges,
a special case of the binary classification tree. Therefore, estimating the threshold points of \eqref{eq1} is equivalent to identify
the splitting points of the tree, which can be achieved by a CART-type procedure \citep{Breiman:etal:1984} described below.
In addition, additivity allows estimating threshold points marginally and provides substantial advantages in practice.

Given observations $\{(X_j^{(i)},y^{(i)})\}_{i=1}^n$ and a subset $ T\subset\mathcal{X}_j$, consider the standard estimator 
$$
\wh p_n( T)=
{\sum_{i=1}^n\mathbb{I}_{\{{y}^{(i)}=1, X_{j}^{(i)}\in  T\}}}\left[{\sum_{i=1}^n\mathbb{I}_{\{X_{j}^{(i)}\in  T\}}}\right]^{-1}, 
$$
by which either the estimated Gini impurity or entropy $\wh \phi_n( T)$ are computed \citep{Breiman:etal:1984}.
First, for each candidate threshold point $t\in T$, define  $T_L = \{x: x\in T,  x< t\}$
and $T_R = \{x: x\in T,  x\ge t\}$ and compute the combined impurity or entropy for the split
$
\wh{\Phi}^n(t, T) = \wh{\phi}_n(T) - \{ \wh{p}_{L,n}(T_L)\wh{\phi}_n(T_L) +
\wh{p}_{R,n}(T_R)\wh{\phi}_n(T_R) \},
$
where $ \wh{p}_{L,n}(T_L)$ and  $\wh{p}_{R,n}(T_R)$ are similarly defined to $\wh p_n(T)$.
Then, estimate $\wh t_{j}$ is obtained by maximizing $\wh \Phi^n(t,T)$ over $T$ for each $j$.
When $K_j\geq 2$, the above steps will be iterated to update $\wh t_{k,j}$ and associated $\{T_L,T_R\}$ until some stopping criterion is reached,
such that $K_j$ threshold points have been obtained.

To establish the non-asymptotic rate of convergence of $\wh t_{j}$'s, notice that for $K_j=1$ model \eqref{eq1} is equivalent to
\begin{equation}\label{model2}
	\text{logit}(p_y) = \beta_0^* + \sum_{j=1}^p \beta_j^* \mathbb{I}_{\{X_j \ge t_j^*\}}.
\end{equation} Denote $p_j^* = \mP_j(X_j \ge t_j^*),  p_{j1}^* = \mP(y=1|X_j \ge
t_j^*),p_{j2}^* = \mP(y=1|X_j < t_j^*)$, and $\phi(x)=-x\log(x)-(1-x)\log(1-x)$ for entropy or  $\phi(x)=2x(1-x)$ for the Gini impurity for $x\in[0,1]$. To establish the uniqueness of the associated variables for solving a relevant optimization problem. We pose the following conditions.

\begin{cond}\label{c3}
	We assume $0 < p_j^* <1$ for each $j$.
\end{cond}

\begin{cond}\label{c4}
	$X_1,\ldots,X_p$ are mutually independent.
\end{cond}

Condition \ref{c3} is regular to exclude the singular and marginal situation of the data.
Condition \ref{c4} is assumed for simplicity of proof, which can be relaxed. And we assume all covariates are continuous
to ensure the uniqueness of the CART-type estimator. Under the above conditions, we have the following theorem.

\begin{theorem}[Uniqueness]
	For continuous covariates $X_1,\ldots,X_p$ in \eqref{model2}, assume Conditions \ref{c3} and \ref{c4} hold. For each $j$, consider the equivalent objective function of the CART-type estimator with the Gini impurity
	$
	\Phi_j(s):= \phi(p_{j1})p_j + \phi(p_{j2})(1-p_j),
	$
	where $\phi(x)=2x(1-x)$, $p_j = \mP_j(X_j \ge s)$, $p_{j1} = \mP(y=1|X_j
	\ge s)$, and $p_{j2} = \mP(y=1|X_j < s)$. Then, for each associated variable $X_j$, the unconstrained optimization problem $\min_{s}  \Phi_j(s)$ admits a unique solution.
	\label{thm1}
\end{theorem}

With $\phi(x)$ replaced by the entropy function in Theorem \ref{thm1}, same conclusion can be drawn. The uniqueness of associated variables guaranteed by Theorem \ref{thm1}, in conjunction with argmax type arguments, along with the ECP (end-cut preference) property of CART \citep{Ishwaran:2015} for non-associated variables, leads to asymptotical behaviors of the CART-type estimator of threshold points.





\begin{theorem}[Asymptotics] Under the conditions in Theorem \ref{thm1},
	\begin{enumerate}
		\item\ For each associated variable $X_j$, the CART-type estimator of the threshold point converges in probability. That is,
		$\wh{t}_{j} \stackrel{p}{\longrightarrow} t_j^*$.
		\item\ For each non-associated variable $X_j$, we have $\mathbb{P}(\wh{t}_{j}\in [t_{j,\min}, t_{j,\min}+\delta] \text{ or } \wh{t}_{j}\in [t_{j,\max}-\delta, t_{j,\max}]) \to  1$, as $n\to \infty$, for some fixed $\delta>0$, where $t_{j,\min}=\max\{\inf\{x: x\in \mathcal{X}_j\}, -n\}$, $t_{j,\max}=\min\{\sup\{x: x\in \mathcal{X}_j\}, n\}$, $\mathcal{X}_j$ is the range of $X_j$ and $n$ is the sample size.
	\end{enumerate}
	\label{thm2}
\end{theorem}

In the Theorem \ref{thm2}, the quantity $n$ in $t_{j,\min}$ and $t_{j,\max}$ can be replaced by any quantity tending to infinity as $n\to \infty$. And $\delta$ in the theorem should be small.
Theorem \ref{thm2} guarantees the consistency for threshold points of associated variables, and for non-associated variables, CART-type estimators will go to either direction of the extreme values of the range. 

The above theorem describes the asymptotic behaviors of associated and non-associated variables, but it doesn't provide the non-asymptotic rate of convergence. Such a rate is indispensable to control the prediction error in Theorem \ref{thm4}, which  quantifies the probability of mistakenly discretizing the covariates using inaccurately estimated threshold points. 
To study the non-asymptotics rate of threshold points, we impose the following conditions.

\begin{cond}\label{c7}
	For each associated variable $X_j$, sample size $n \ge \left({2}/{c_j}\right)^{(\delta_0-\delta)^{-1}}$, where constant $c_j>0$ depends on $t_j^*$ only and $\delta_0,\delta$ satisfy $1/2> \delta_0> \delta>0$.
\end{cond}

\begin{cond}\label{c8}
	For each associated variable $X_j$, the distribution of $X_j$  has nonzero density at $t_j^*$.
\end{cond}

Condition \ref{c7} gives the necessary sample size to reach the desired convergence rate, and Condition \ref{c8} ensures that there is some information near the true threshold points of associated variables. With these conditions, we   have the following result.

\begin{theorem}[Non-asymptotic rate of convergence for associated variables] Assume  conditions in Theorem \ref{thm2} hold, and for associated variables, Conditions \ref{c7} and \ref{c8} hold. Then, with probability at least $1 - 2\exp \left(-n^{2\delta}/8\right)$, $|\wh{t}_{j} - t_j^*| \le n^{-{1}/{2}+\delta_0}$, for each associated variable $X_j$.
	\label{thm3}
\end{theorem}

Theorem \ref{thm3} suggests that, due to its nonparametric nature, finite sample convergence rate of the proposed CART-type estimator $\wh{t}_{j}$ for associated variables is slightly slower than $n^{-1/2}$. On the other hand, exponential tail provided by Theorem \ref{thm3} facilitates establishing selection consistency of FILTER.
Here we do not show the non-asymptotic behavior for non-associated variables. As shown in Theorem \ref{thm4}, no matter what the estimated threshold points for non-associated variables are, the results in the previous section always hold. 

With above discussion, we can conclude that CART-type estimators for the threshold points in Model \eqref{eq1} can satisfy the conditions in Theorem \ref{thm4}. In Section \ref{simulation.sec}, we provide an estimation strategy for threshold points, which shows satisfactory performance in practice. 



\section{Monte Carlo Evidences}
\label{simulation.sec}

\subsection{Simulation design}

In this section, we compare the FILTER with peer competitors via Monte Carlo simulation studies. We consider various settings; see blow for details. In all settings, we always consider covariates being a $p$-dimensional normal random vector $\bX\sim \mathcal{N}(\boldsymbol{0}, \bSigma)$, where $\bSigma \in \mathbb{R}^{p \times p}$ is the Auto-Regression correlation matrix with the $(i,j)$-element of $\bSigma$ being $\rho^{|i-j|}$ for a given $\rho$. Denote $\rho=0$ for the mutual independent case among covariates. Moreover, such a setting of covariates satisfies the condition \ref{c13} as shown in the supplementary materials.

\subsubsection{Estimation and selection}\label{sec311}

To illustrate the estimation performance for threshold points and the variable selection capability of the FILTER, we consider the model \eqref{model2} with one threshold point for each covariate. 

For such a model, we consider settings with sample sizes $n=100, 150, \ldots, 400$, dimension $p=500$ for covariates and $p_0=5$ as the sparsity level. For covariates, we set $\rho=0$ in $\bSigma$ as required by the condition \ref{c4}. We set the true threshold points $t_{j}^{*}=0$ for $j=1,\ldots, p$, and take the true regression coefficients $\beta_{j}^{*}=3$ for $j=1,\ldots,p_0$ and $\beta_{j}^{*}=0$ for $j>p_{0}$. 
To balance the positive and negative class in the data, we set the intercept $\beta_0^{*} = - n^{-1} \sum_{i=1}^n \sum_{j=1}^{p}\beta_j^{*} \mathbb{I}_{\{X_j \ge t_j^*\}}$. 
Finally, the response $y$ is generated from the model \eqref{model2}. 
For each setting, we independently generate $500$ replications. 


To estimate threshold points, we employ a Bagging technique in conjunction with the proposed CART-type estimator. Specifically, for a given dataset, we first randomly select sample points with replacement to form a Bagging dataset. Secondly, we perform the CART-type procedure described in Section \ref{sec22} on the Bagging dataset to has an intermediate estimator of the threshold point for each covariate. This procedure is then repeated $100$ times independently, and the average of these intermediate estimators for each covariate is used as the final estimated threshold point correspondingly. 
Based on these estimates, the FILTER model was fitted by solving \eqref{transloss}, where $\lambda$ is chosen using a $5$-folds cross validation.

\subsubsection{Prediction}\label{sec312}

To illustrate the prediction performance of the FILTER with peer competitors, we consider two families of models. That is, (I) the FILTER model \eqref{eq01} with $K_{j}=3, j=1, \ldots, p$, and (II) a model related to the one in \cite{Friedman:1991}, who is termed the piecewise model in this paper, with the probability of success $p_y=\mP(y=1|\bX)$ satisfying $\text{logit}(p_y) = \beta_0^* + \sum_{j=1}^{p} h(X_j)$, 
where $  h(X_j)  = \beta_{0,j}^* \mathbb{I}_{\{X_j\le t_{1,j}^*\}}
	+ \beta_{1,j}^*\sin(\pi X_j) \mathbb{I}_{\{t_{1,j}^*< X_j \le t_{2,j}^*\}} 
 + \beta_{2,j}^* (X_j-0.5)^2  \mathbb{I}_{\{t_{2,j}^*< X_j \le t_{3,j}^*\}}
	+\beta_{3,j}^* X_j  \mathbb{I}_{\{t_{3,j}^*< X_j\}}$.
Such a piecewise model is applied to evaluate the performance of the proposed estimator in the presence of model misspecifications.

For both families of models, we consider settings with sample sizes $n=400, 800$, dimension of covariates $p=500$ and sparsity level $p_0=5$.
For covariates, we set $\rho=0, 0.5$ in $\bSigma$ representing the independent case and the correlated case respectively.
We set the true threshold points $t_{k,j}^{*}=\Phi^{-1}((1+k)/6)$ for $k=1,2,3, j=1,\ldots, p$ with $\Phi^{-1}(\cdot)$ being the quantile function of the standard normal distribution. 
Regarding the true regression coefficients, we take $\beta_{0,j}^*=0$, $\beta_{1,j}^*=\beta_{3,j}^*=5$ and $\beta_{2,j}^*=10$ for $j=1,\ldots,p_0$; $\beta_{k,j}^*=0$ otherwise. 
We set the intercept such that the positive and negative class are balanced in the data as in section \ref{sec311}.
Finally, the response $y$ is generated from the aforementioned two families of models respectively. 
For each setting, we independently generate $500$ replications. In each replication, $80\%$ of data points are used for training the model, and $20\%$ of data points are used for testing the trained model.

The estimation procedure of the FILTER for the two families of models is similar to the one in section \ref{sec311} with some modifications for the estimation of threshold points. To estimate threshold points, instead of taking average of intermediate estimators obtained by the CART-type procedure on the Bagging dataset for each covariate, we employed $K$-means for these intermediate estimators to form $K$ estimated threshold points for each covariate. For all settings, we set $K=6$. 

For peer competitors, we consider CART, random forest (RF), CART and Bagging (CB), $\ell_1$-regularized logistic regression ($\ell_1$-logistic), and refitted $\ell_1$-regularized logistic regression (logistic-refit). 
For the CB method, the prediction of a given test sample point is based on the estimation procedure of threshold points mentioned above. Specifically, we obtain an intermediate predicted probability of success for the test sample point using the CART model fitted on each Bagging dataset first. The average of these intermediate probabilities for the test sample point is then considered as its predicted probability of success.
For the logistic-refit, we fit a logistic regression based on the variable selected by $\ell_1$-penalty. 


\subsection{Results}\label{sec:simu-results}

\paragraph{Estimation and selection}

For the performance of threshold points estimation, we consider the mean absolute bias, denoted as $\text{MAB}_{\text{t}}$. This criterion is defined as $ \text{MAB}_{\text{t}}  = {|\text{TS}|}^{-1}\sum_{j\in \text{TS}} |\wh{t}_j - t^*_j|$, where $\text{TS}$ is the support of the true regression coefficients vector.
Besides, the root squared error ($\text{RSE}_{\text{est}}$) and the root mean squared error ($\text{RMSE}_{\text{est}}$) are used as criteria for the performance of regression coefficient estimation. Conventionally, the $\text{RSE}_{\text{est}}$ and $\text{RMSE}_{\text{est}}$ of an estimator $\wh \bbeta$ for the true regression coefficients vector $\bbeta^* \in \mathbb{R}^{p}$ are defined by $ \|\wh{\bbeta} - \bbeta^*\|_2 $ and $p^{-1/2}\|\wh{\bbeta} - \bbeta^*\|_2$, respectively. 
Lastly, to evaluate the capability of variable selection, we consider sensitivity and specificity, denoted as $\text{SEN}_{\text{vs}}$ and $\text{SPE}_{\text{vs}}$ and defined as
$\text{SEN}_{\text{vs}}={\mathrm{TP}_{\text{vs}}}(\mathrm{TP}_{\text{vs}}+\mathrm{FN}_{\text{vs}})^{-1}$ and $\text{SPE}_{\text{vs}}={\mathrm{TN}_{\text{vs}}}(\mathrm{TN}_{\text{vs}}+\mathrm{FP}_{\text{vs}})^{-1}$ respectively, where $\mathrm{TP}_{\text{vs}}$, $\mathrm{FP}_{\text{vs}}$,  $\mathrm{FN}_{\text{vs}}$, and $\mathrm{TN}_{\text{vs}}$ are the number of correctly selected nonzero coefficients, falsely selected zero coefficients, falsely excluded nonzero coefficients, and correctly excluded zero coefficients, respectively.
The results for settings in section \ref{sec311} are listed in Table \ref{tab:vs}. 

\begin{figure}[bt]
	\centering
	\subfigure{
	\includegraphics[width=0.425\textwidth]{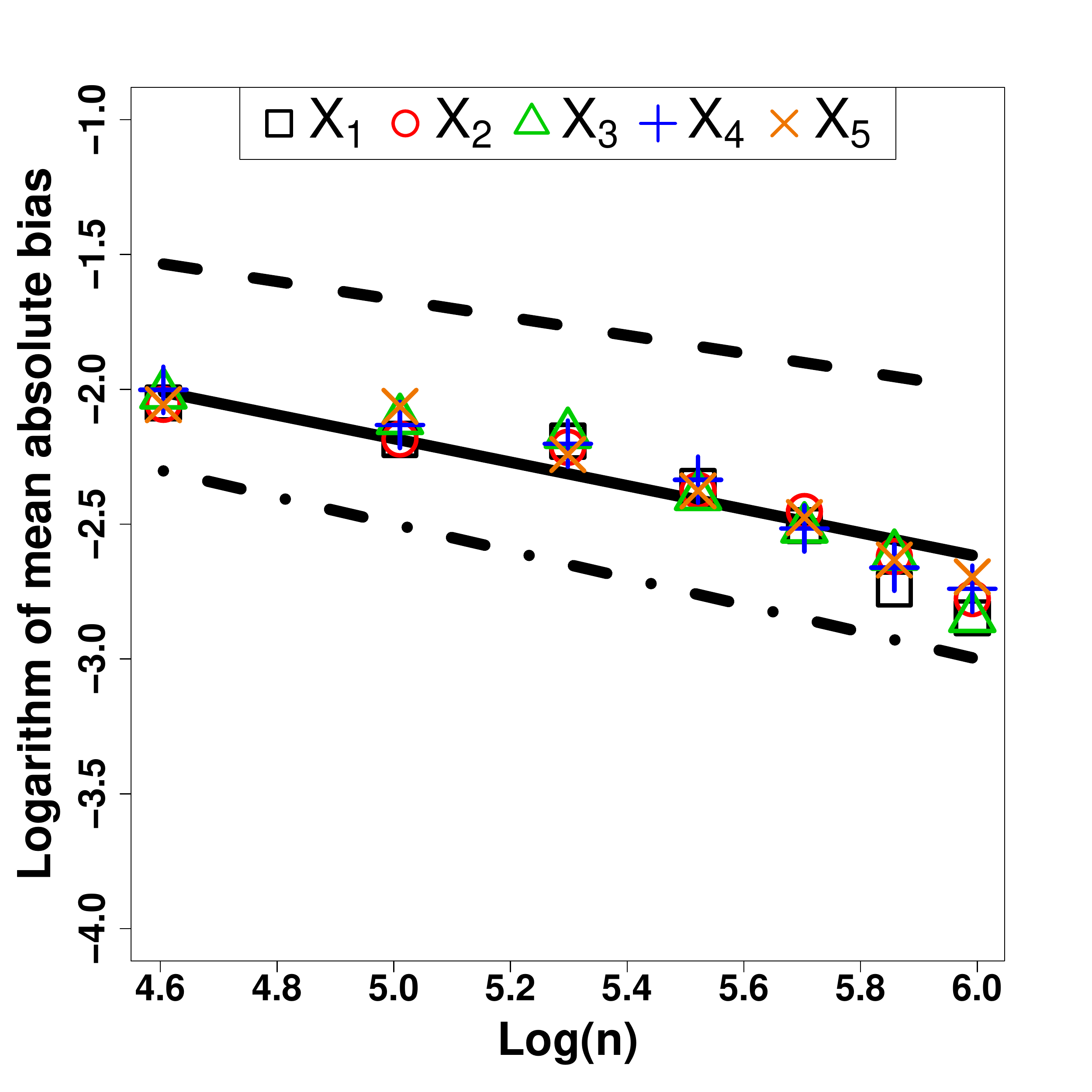}
	}
	\subfigure{
		\includegraphics[width=0.425\textwidth]{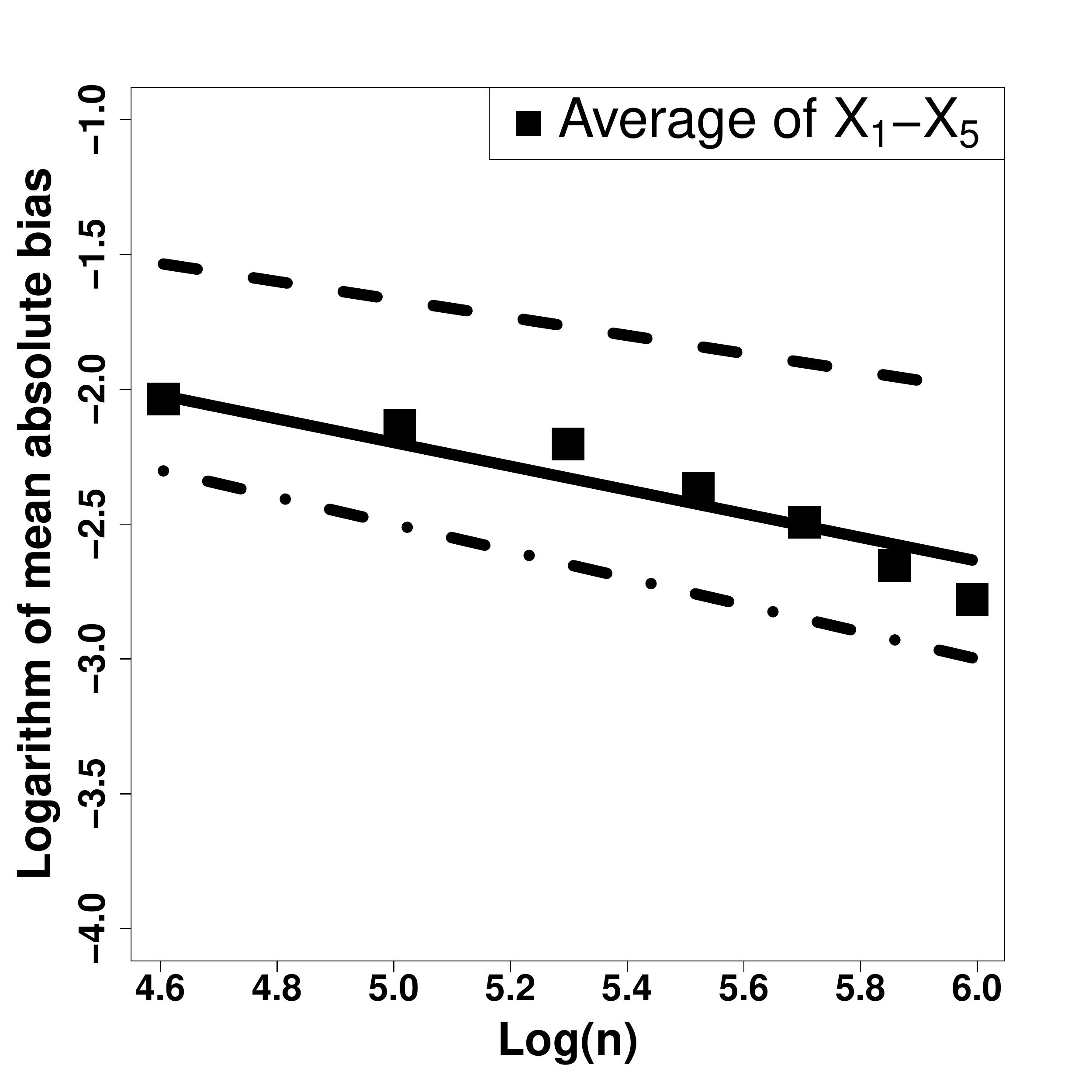}
	}  
	\caption{Plots of $\log(\text{MAB}_{\text{t}})$  against $\log n$ with $n=100,150,\ldots,400$. The left panel shows results for the individual estimated threshold points of the $p_0$ nonzero variables, and the right panel shows results for averages of the estimated threshold points of the $p_0$ nonzero variables. The solid lines are simple linear regressions based on simulations, and the dot-dash and dashed lines represent $y=-x/2$ and $y=-x/3$, respectively. Results are based on $500$ replications.}
	\label{mb_cp}
\end{figure}

Table \ref{tab:vs} shows that, for our method,  both mean absolute bias for estimated threshold points and errors for the estimated regression coefficients decrease as $n$ growing, and sensitivity and specificity for variable selection increase in $n$. 
These provide empirical evidences confirming Theorem \ref{thm4}. In addition, Figure \ref{mb_cp} displays  $\log(\text{MAB}_{\text{t}})$  against $\log n$ for our CART-type estimator. The negative regression slope is fairly close to $1/2$ and larger than $1/3$, which validates Theorem \ref{thm3} as expected. In summary, the proposed estimator on the FILTER performs reasonably well and agrees with the theoretical guarantees.

\begin{table}[bt]
	\centering
	\caption{Empirical performance on estimated threshold points and regression coefficients for FILTER under settings in Section \protect\ref{sec311}. For each measure, averages with standard deviations (in the parentheses) from $500$ replications under different settings are given.}
	\begin{threeparttable}
		\begin{tabular}{cccccc}
		    \toprule
			\textbf{n} & \textbf{$\text{MAB}_{\text{t}}$} & \textbf{$\text{RSE}_{\text{est}}$} & \textbf{$\text{RMSE}_{\text{est}}$} & \textbf{$\text{SEN}_{\text{vs}}$} & \textbf{$\text{SPE}_{\text{vs}}$}\\
		    \midrule
			100   & 0.13(0.19) & 5.92(0.69) & 0.26(0.03) & 0.80(0.23) & 0.95(0.04) \\
			150   & 0.12(0.19) & 5.33(0.55) & 0.24(0.02) & 0.94(0.13) & 0.95(0.05) \\
			200   & 0.11(0.18) & 4.98(0.48) & 0.22(0.02) & 0.98(0.06) & 0.96(0.05) \\
			250   & 0.09(0.16) & 4.75(0.44) & 0.21(0.02) & 1.00(0.02) & 0.96(0.04) \\
			300   & 0.08(0.14) & 4.64(0.45) & 0.21(0.02) & 1.00(0.01) & 0.97(0.03) \\
			350   & 0.07(0.12) & 4.44(0.47) & 0.20(0.02) & 1.00(0.01) & 0.97(0.03) \\
			400   & 0.06(0.11) & 4.30(0.52) & 0.19(0.02) & 1.00(0.00) & 0.97(0.03) \\
			\hline  
		\end{tabular}
	\end{threeparttable}
	\label{tab:vs}
\end{table}

\paragraph{Prediction}

To compare the performance of our proposal on predictions with aforementioned peer competitors in Section \ref{sec312}, we consider area under the curve (AUC) and three strictly proper scoring rules, namely logarithmic score (Logs), continuous ranked probability score (CRPS) and Brier score (Brier), as criteria. 
Definitions of these scoring rules can be found in \cite{GR:2007}. 
AUC is a widely used measure which summarizes the ability of a classifier to distinguish between classes, and proper scoring rules encourage the forecaster to make careful assessments and to be honest \citep{GR:2007}, which are more fair to compare the predicted probabilities of different methods. \texttt{R} packages \texttt{rpart}, \texttt{AUC} and \texttt{scoringRules} are used to compute these criteria.


Results for families (I) and (II) are displayed in Tables \ref{scn2thr} and \ref{scn2piece}, respectively.
For the family (I), given a small sample size, our proposal and CB enjoy comparable performance and are better than others in general, while our model provides easier interpretation. As the sample size increasing, our method outperforms all the other methods. 
Interestingly, though the theoretical guarantees are based on mutual independence of $X_j$'s, numerical studies show that our method works reasonably well in the presence of correlations among covariates. 
For the family (II), similar observations are made. The codes for FILTER and simulation models can be found at \url{https://github.com/ynlin11/FILTER }.

\begin{table}[bt]
	\centering
	\caption{Empirical performances on predictions for the family (I) of our proposal in comparison to peer competitors. For each measure, averages with standard deviations (in the parentheses) from $500$ replications under different settings are given. The methods with highest averages are marked in bold.}
	\scalebox{0.85}{\footnotesize
	\begin{threeparttable}
		\begin{tabular}{ccccccc}
			\toprule
			\textbf{$\rho$} & \textbf{n} & \textbf{Method} & \textbf{AUC} & \textbf{Logs} & \textbf{CRPS}  & \textbf{Brier} \\
		    \midrule
            \multirow{12}{*}{0} & \multirow{6}{*}{200} & CB    & \textbf{0.75(0.09)} & \textbf{-0.62(0.05)} & \textbf{-0.21(0.02)} & \textbf{-0.43(0.05)} \\
                  &       & CART  & 0.65(0.09) & -0.85(0.22) & -0.28(0.07) & -0.55(0.13) \\
                  &       & RF    & 0.68(0.09) & -0.67(0.02) & -0.24(0.01) & -0.48(0.02) \\
                  &       & $\ell_1$-logistic & 0.61(0.10) & -0.78(0.26) & -0.26(0.05) & -0.52(0.10) \\
                  &       & logistic-refit & 0.60(0.10) & -4.45(5.07) & -0.34(0.10) & -0.68(0.21) \\
                  &       & FILTER & \textbf{0.75(0.09)} & -0.65(0.15) & -0.22(0.04) & -0.44(0.08) \\
        \cmidrule{2-7}          & \multirow{6}{*}{400} & CB    & 0.80(0.05) & -0.56(0.04) & -0.19(0.02) & -0.38(0.04) \\
                  &       & CART  & 0.68(0.07) & -0.80(0.16) & -0.26(0.05) & -0.51(0.10) \\
                  &       & RF    & 0.77(0.06) & -0.65(0.01) & -0.23(0.01) & -0.46(0.01) \\
                  &       & $\ell_1$-logistic & 0.69(0.07) & -0.66(0.13) & -0.23(0.02) & -0.46(0.04) \\
                  &       & logistic-refit & 0.67(0.08) & -1.63(3.21) & -0.26(0.07) & -0.52(0.13) \\
                  &       & FILTER & \textbf{0.84(0.05)} & \textbf{-0.53(0.06)} & \textbf{-0.18(0.02)} & \textbf{-0.35(0.04)} \\
        \cmidrule{1-7}    \multirow{12}{*}{0.5} & \multirow{6}{*}{200} & CB    & 0.86(0.07) & \textbf{-0.49(0.07)} & \textbf{-0.16(0.03)} & \textbf{-0.32(0.06)} \\
                  &       & CART  & 0.75(0.09) & -0.66(0.19) & -0.21(0.06) & -0.42(0.12) \\
                  &       & RF    & 0.85(0.07) & -0.61(0.03) & -0.21(0.01) & -0.42(0.03) \\
                  &       & $\ell_1$-logistic & 0.80(0.08) & -0.59(0.08) & -0.20(0.03) & -0.39(0.06) \\
                  &       & logistic-refit & 0.78(0.09) & -1.33(2.40) & -0.21(0.07) & -0.42(0.14) \\
                  &       & FILTER & \textbf{0.87(0.07)} & -0.50(0.10) & \textbf{-0.16(0.04)} & \textbf{-0.32(0.07)} \\
        \cmidrule{2-7}          & \multirow{6}{*}{400} & CB    & 0.88(0.04) & -0.45(0.05) & -0.15(0.02) & -0.29(0.04) \\
                  &       & CART  & 0.78(0.06) & -0.62(0.13) & -0.19(0.04) & -0.38(0.08) \\
                  &       & RF    & 0.88(0.04) & -0.58(0.02) & -0.20(0.01) & -0.39(0.02) \\
                  &       & $\ell_1$-logistic & 0.83(0.05) & -0.54(0.05) & -0.18(0.02) & -0.36(0.04) \\
                  &       & logistic-refit & 0.81(0.06) & -0.72(1.21) & -0.18(0.04) & -0.37(0.08) \\
                  &       & FILTER & \textbf{0.90(0.04)} & \textbf{-0.43(0.06)} & \textbf{-0.14(0.02)} & \textbf{-0.27(0.04)} \\
			\hline  
		\end{tabular}
	\end{threeparttable}
	}
	\label{scn2thr}
\end{table}

\begin{table}[bt]
	\centering
	\caption{Empirical performances on predictions for the family (II) of our proposal in comparison to peer competitors. For each measure, averages with standard deviations (in the parentheses) from $500$ replications under different settings are given. The methods with highest averages are marked in bold.}
	\scalebox{0.85}{\footnotesize
	\begin{threeparttable}
		\begin{tabular}{ccccccc}
			\toprule
			\textbf{$\rho$} & \textbf{n} & \textbf{Method} & \textbf{AUC} & \textbf{Logs} & \textbf{CRPS}  & \textbf{Brier} \\
			\midrule
            \multirow{12}{*}{0} & \multirow{6}{*}{200} & CB    & \textbf{0.80(0.08)} & \textbf{-0.58(0.06)} & -0.20(0.03) & -0.39(0.05) \\
                  &       & CART  & 0.69(0.10) & -0.76(0.21) & -0.24(0.07) & -0.49(0.14) \\
                  &       & RF    & 0.72(0.09) & -0.66(0.02) & -0.24(0.01) & -0.47(0.02) \\
                  &       & $\ell_1$-logistic & 0.72(0.10) & -0.67(0.16) & -0.23(0.04) & -0.45(0.08) \\
                  &       & logistic-refit & 0.69(0.10) & -3.65(4.48) & -0.29(0.09) & -0.57(0.18) \\
                  &       & FILTER & \textbf{0.80(0.08)} & \textbf{-0.58(0.12)} & \textbf{-0.19(0.04)} & \textbf{-0.38(0.07)} \\
        \cmidrule{2-7}          & \multirow{6}{*}{400} & CB    & 0.86(0.05) & -0.50(0.05) & -0.16(0.02) & -0.32(0.04) \\
                  &       & CART  & 0.74(0.07) & -0.68(0.15) & -0.22(0.05) & -0.43(0.09) \\
                  &       & RF    & 0.82(0.05) & -0.64(0.01) & -0.22(0.01) & -0.45(0.01) \\
                  &       & $\ell_1$-logistic & 0.80(0.05) & -0.58(0.04) & -0.20(0.02) & -0.39(0.03) \\
                  &       & logistic-refit & 0.78(0.07) & -0.87(1.58) & -0.21(0.05) & -0.42(0.10) \\
                  &       & FILTER & \textbf{0.88(0.04)} & \textbf{-0.48(0.05)} & \textbf{-0.15(0.02)} & \textbf{-0.31(0.04)} \\
          \cmidrule{1-7}  \multirow{12}{*}{0.5} & \multirow{6}{*}{200} & CB    & 0.91(0.05) & -0.42(0.07) & \textbf{-0.13(0.03)} & -0.26(0.05) \\
                  &       & CART  & 0.79(0.08) & -0.56(0.18) & -0.17(0.06) & -0.35(0.11) \\
                  &       & RF    & 0.90(0.05) & -0.58(0.03) & -0.20(0.01) & -0.39(0.03) \\
                  &       & $\ell_1$-logistic & 0.88(0.06) & -0.47(0.08) & -0.15(0.03) & -0.30(0.05) \\
                  &       & logistic-refit & 0.85(0.08) & -1.88(2.97) & -0.17(0.08) & -0.35(0.15) \\
                  &       & FILTER & \textbf{0.92(0.05)} & \textbf{-0.41(0.08)} & \textbf{-0.13(0.03)} & \textbf{-0.25(0.06)} \\
        \cmidrule{2-7}          & \multirow{6}{*}{400} & CB    & 0.93(0.03) & -0.37(0.05) & -0.11(0.02) & -0.23(0.04) \\
                  &       & CART  & 0.83(0.06) & -0.51(0.13) & -0.16(0.04) & -0.31(0.08) \\
                  &       & RF    & 0.93(0.03) & -0.54(0.02) & -0.18(0.01) & -0.35(0.02) \\
                  &       & $\ell_1$-logistic & 0.91(0.04) & -0.42(0.05) & -0.13(0.02) & -0.26(0.04) \\
                  &       & logistic-refit & 0.89(0.05) & -0.81(1.69) & -0.14(0.05) & -0.27(0.09) \\
                  &       & FILTER & \textbf{0.94(0.03)} & \textbf{-0.34(0.05)} & \textbf{-0.10(0.02)} & \textbf{-0.21(0.04)} \\
			\hline  
		\end{tabular}
	\end{threeparttable}
	}
	\label{scn2piece}
\end{table}

\section{Application of FILTER on Diabetes Prediction}
\label{real}

In this section, we apply the FILTER, as well as some competitive methods for comparisons, on a real data for diabetes prediction. Moreover, we shall develop a risk score using the FILTER model trained on the data.
The data considered here is the annual physical examination/survey data collected from research institutes in Beijing, China \citep{Plos}. 
In the data, the participants, aged $20$ and above, are educated and in a sedentary working pattern. Forty-three factors are measured and recorded, such as body mass index (BMI), diastolic and systolic blood pressures (DBP/SBP), high density lipoprotein cholesterol (HLD-C), etc. A list of factors is included in the supplementary materials. The amount of fasting blood sugar is used to define the response. Specifically, a subject is labeled as normal if the fasting blood sugar level is less than $7.0\text{mmol}/\text{L}$ and as diabetes otherwise. Excluding subjects with missing values, a sample of size $n=4,940$ with about $3.32\%$ diagnosed cases remains for analysis. It can be seen the imbalance issue can not be ignored in our data, and more cares must be taken when analyzing it. 

\begin{figure}[h]
 	\centering
 	\includegraphics[scale=0.25]{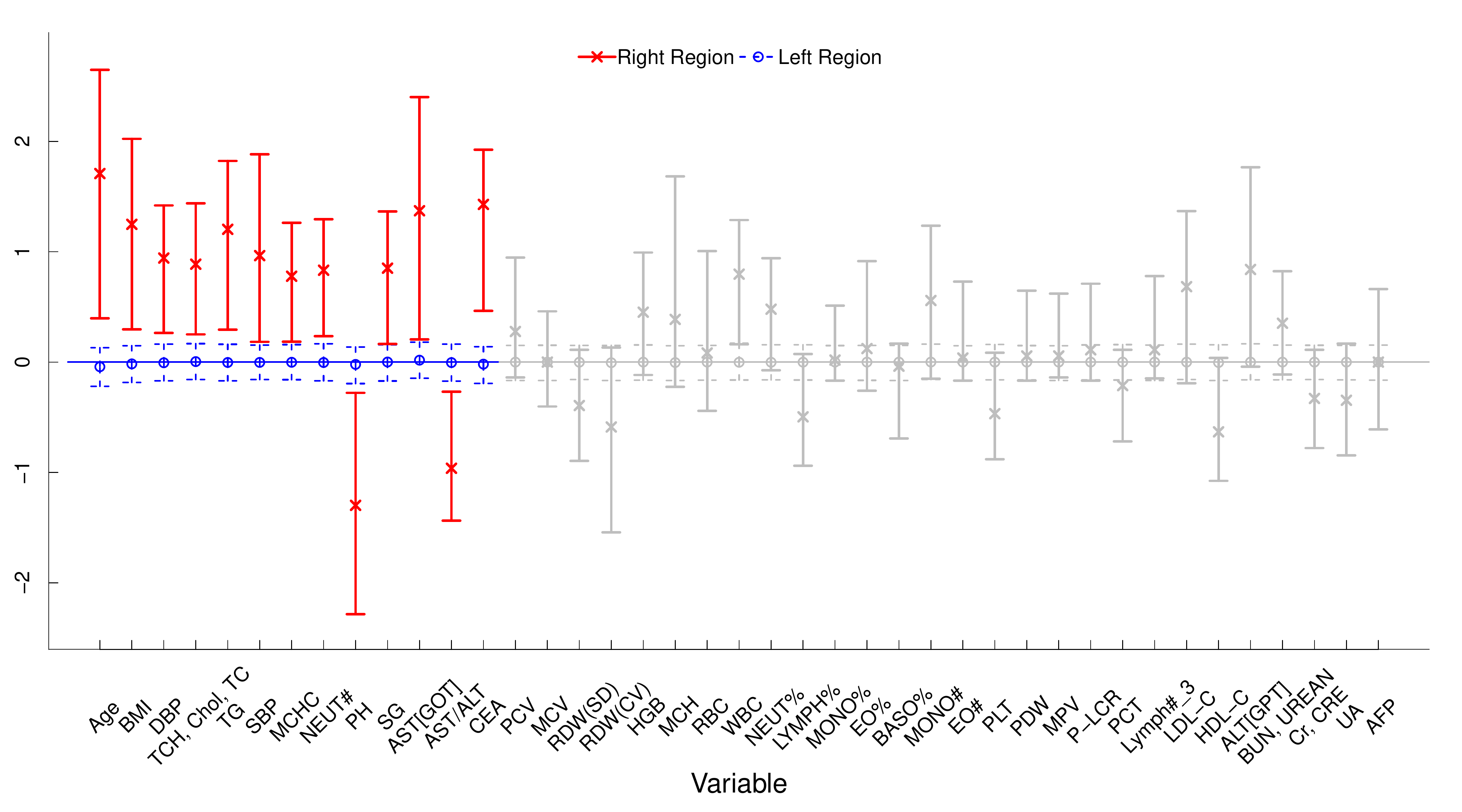}
 	\caption{Threshold effects of covariates in the annual physcial examination data from Beijing \protect\citep{Plos}, based on linear threshold model from Banerjee and McKeague (2007).}\label{fig:ci}
 \end{figure}

As a pre-analysis, a non-standard cubic-root consistent estimators based confidence intervals have been developed \citep{Banerjee:McKeague:2007} on the covariates in our physical examination data from Beijing; see in Figure \ref{fig:ci}.
These confidence intervals give us confidence on the existence of threshold effect among the covariates.

\subsection{Prediction Performances}
\label{GOF}

In this subsection, we report prediction performances of the FILTER and peer competitors, including CART, RF, CB, $\ell_1$-logistic, and logistic-refit as discussed in Section \ref{sec312}, on the diabetes dataset from two aspects. One is summary measures including AUC and partial AUC, and the other one is the Murphy diagram, a tool permitting detailed comparisons of forecasting methods \citep{Ehm2016}. 

First, we consider summary measures to evaluate the prediction performance. To this end, we perform a $5$-fold cross-validation analysis. Due to the imbalance issue in our data, AUC and partial AUC (pAUC), computed by \texttt{R} packages \texttt{AUC} and \texttt{pROC} respectively, are adopted as measures of the prediction performance.
For the data with the imbalance issue, we need to pay more attention to the low false positive rate (FPR) region on the receiver operating characteristic (ROC) curve. In that case, partial AUC, originally introduced in \cite{McClish:1989} and defined as a portion of the FPR, is a reasonable measure. Additionally, partial AUC can be standardized with the following formula \citep{Robin:etal:2011}:
\[
\frac{1}{2} \left( 1 + \frac{\mathrm{pAUC}-\min}{\max-\min} \right),
\]
where $\mathrm{pAUC}$ is the partial AUC over the specific region, $\min$ is the partial AUC over the same region of the diagonal ROC curve, and $\max$ is the partial AUC over the same region of the perfect ROC curve. The result is a standardized partial AUC which is always 1 for a perfect ROC curve and 0.5 for a non-discriminant ROC curve. 
To compute partial AUC, we consider the region with FPR ranging from $0$ to $0.1$ in our analysis.

In our analysis, the prediction of a test sample point is made based on whether or not the predicted probability of diabetes exceeds the marginal proportion of diabetes in the original data. Such a cut-off has been used for case-control sampling in the presence of imbalance of classes \citep{PP79}, and other methods could be used in practice as well \citep{FH14}.
For the FILTER, we consider the estimation procedure in section \ref{sec312}, and set the number of clusters $K$ with $K=2,4,6,8$ in the $K$-means step for estimating threshold points.
Implied by results in Table \ref{realdata}, the FILTER is fairly robust for the choice of $K$ while it performs better than all the methods on prediction most of the time (especially when $K=6$).

\begin{table}[bt]
	\centering
	\caption{Comparison of the performance on diabetes prediction of the proposed method with different choices of $K_j$'s to that of existing methods using the physical examination/survey data collected from research institutes in Beijing, China.}
	\begin{threeparttable}
		\begin{tabular}{lccc}
			\toprule
			\textbf{Method} & \textbf{AUC} & \textbf{pAUC} & \textbf{pAUC(standardized)} \\
			\midrule
			CART  & 0.7214 & 0.0227 & 0.5930 \\
			RF    & 0.8413 & 0.0358 & 0.6622 \\
			CB    & 0.8325 & 0.0334 & 0.6492 \\
			logistic-refit & 0.8624 & 0.0402 & 0.6853 \\
			$\ell_1$-logistic & 0.8623 & 0.0386 & 0.6771 \\
			FILTER(K=2) & 0.8427 & 0.0355 & 0.6605 \\
			FILTER(K=4) & 0.8642 & 0.0406 & 0.6875 \\
			FILTER(K=6) & 0.8682 & 0.0411 & 0.6903 \\
			FILTER(K=8) & 0.8685 & 0.0387 & 0.6774 \\
			\hline  
		\end{tabular}
	\end{threeparttable}
	\label{realdata}
\end{table}

For more detailed comparisons, we consider Murphy diagrams proposed by \cite{Ehm2016}. In that paper, the authors showed that every scoring function consistent for the expectile can be represented as a mixture of extremal scoring functions. In binary classification case, extremal scoring function is the distance of $y$ from the true probability of success $p$ if the predicted probability of success $\hat{p}$ is on the opposite side from $y$, and otherwise it is zero. Therefore, the smaller the extremal scoring functions are, the better. An extremal scoring function can be further controlled by a parameter $\alpha\in (0,1)$ called level. When $y$ is smaller than $p$, the extremal scoring function should be weighted by $1-\alpha$, otherwise by $\alpha$. Hence, the authors proposed Murphy diagrams as averages of empirical extremal scoring functions. In summary, the Murphy diagram is a way to graphically assess the predicted probabilities of a method with varying thresholds for the probability of success. 

We use $5$-folds cross-validation analysis with Murphy diagrams as criterion to compare different methods. In a Murphy diagram, each curve of the empirical score is plotted pointwisely for the parameter $p$. Therefore, for each value of $p$, we take the average of the predicted probabilities of $5$ folds as the overall prediction of $p$. 
Due to the imbalance issue, we set level $\alpha=0.9$ when plotting Murphy diagrams. Such a choice would share more weight on positive cases, who are more crucial for the analysis. The computations are conducted by the \texttt{R} package \texttt{murphydiagram}. 

\begin{figure}[bt]
	\centering
	\subfigure[Murphy diagrams for different methods]{
		\includegraphics[width=0.475\textwidth]{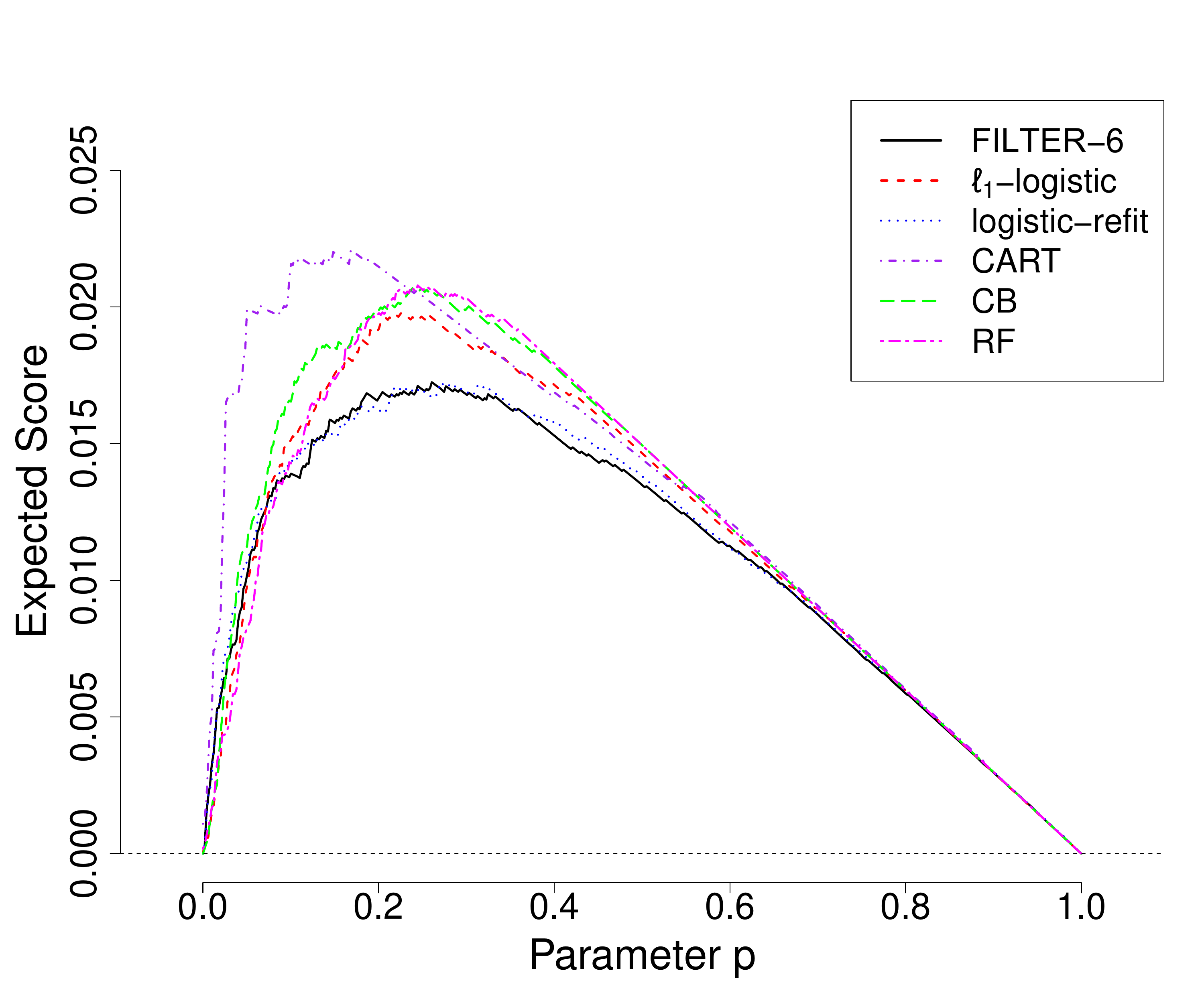}
	}
	\subfigure[Best forecast method comparison]{
		\includegraphics[width=0.475\textwidth]{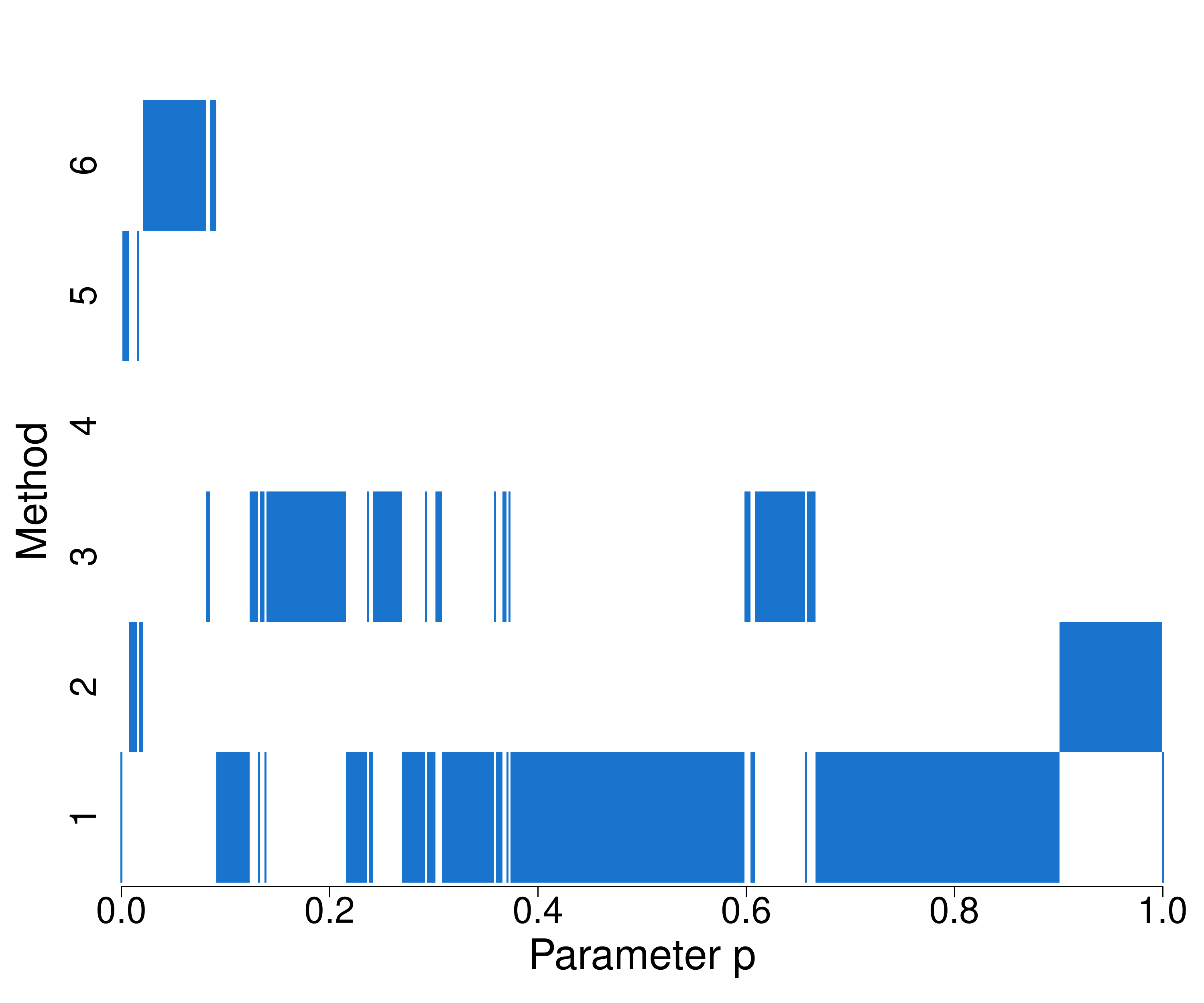}
	}
	\caption{The left and right panels are Murphy diagrams for different methods and the best forecast method for varying parameter $p$ on the physical examination/survey data from Beijing, China for diabetes prediction in Section \protect\ref{real}. In (b), 1:FILTER-$6$, 2:$\ell_1$-logistic, 3:logistic-refit, 4:CART, 5:CB, 6:RF.
	}\label{Fig1}
\end{figure}

In Figure \ref{Fig1}, the left panel shows comparisons of the FILTER with $K=6$ (FILTER-$6$) and other aforementioned methods, and the right panel indicates the method with the lowest empirical score for varying $p$. As we can see, the FILTER has the lowest empirical scores most of the time, and it is comparable with the lowest one otherwise. This shows the relative merits on prediction of the FILTER model. More supported evidences can be found in the supplementary materials. 



\subsection{FILTER-based risk score}

In this subsection, we sketch the construction of FILTER-based risk score using $\wh \beta_{k,j}$. For each covariate $\hat Z_j$, since the $K_j$ coefficients may be fused into a fewer number by the algorithm, we denote the resulted coefficients with a prime, and denote the combined coefficients as $\wh \bbeta_j^{\text{comb}}=(\wh \beta_{1,j}',\ldots, \wh \beta_{K_j',j}')^{\top}$. This can guarantee $\wh \beta_{k,j}' \neq \wh \beta_{k+1,j}'$ for each $k=1,\ldots, K_j'-1$. 
Secondly, to guarantee the coefficient of the level with lowest value being zero, let $\wh \bbeta_j^{\sharp}=\wh \bbeta_j^{\text{comb}} - \alpha_j\textbf{1}$, where $\alpha_j=\min_{1\leq k\leq K_j'} \wh \beta_{k,j}'$ and $\textbf{1}\in \mathbb{R}^{K_j'}$. 
Next, compute $\widetilde{\bbeta}_j=100\wh \bbeta_j^{\sharp}
(\sum_{j=1}^p||\wh \bbeta_j^{\sharp}||_{\infty})^{-1}$. 
Finally, for each individual, the risk score is defined by $\text{RS}^{\text{FILTER}}=\sum_{j=1}^p \widetilde{\beta}_{k_{ij},j}$ where $k_{ij}\in\{1,\ldots, K_j'\}$ denotes the level of the $j$th factor of individual $i$. 
Thus, the minimum score   $0$   corresponds to the
lowest risk of diabetes while the subject with all factors reaching the highest level will have a score $100$ indicating a high risk. The $\text{RS}^{\text{FILTER}}$ inherits the flexibility and reliable prediction from the FILTER model, as well as easy interpretation.

To construct the FILTER-based risk score, we need to decide the parameter $K$, the number of clusters, in the $K$-means step for estimating threshold points. Thereby, we select $K$ ranging from $2$ to $10$ by the cross-validation with AUC as criterion. Moreover, the two standard error rule is used in our analysis for the simpler model with more interpretation in the cross-validation procedure. Such an analysis results $K$ to be $6$.

For the FILTER with $K=6$, thirty five factors are selected (see the discussion after Theorem \ref{thm4}), where contributions to the risk score of sixteen factors with more than 3 levels are displayed in Table \ref{score15}. We obtain the FILTER-based risk scores for all subjects within the data. The median and mean scores are $14.88$ and $14.98$ reflecting the symmetry of the score distribution, the upper $5\%$ and $1\%$ quantiles of scores are $22.65$ and $25.9$, and the minimum and maximum scores are $0.96$ and $32.52$, respectively. The $3.32\%$ upper quantile of scores is $23.3$, which can serve as the critical point for diabetes prognosis.
More importantly, age, BMI, DBP, Triglyceride (TG), total cholesterol (TC) and a few lab testing indices are considered by the FILTER to be diabetes relevant. They provide much clearer clinical interpretations.


\begin{table}[bt]
	\centering
	\caption{Diabetes risk scores contributed by $16$ factors with levels more than $3$ selected by the FILTER with $K=6$ using the physical examination/survey data from Beijing, China. The abbreviations are annotated in Table \protect\ref{abbr} in the supplementary materials.}
	{\footnotesize
	\begin{threeparttable}
		\begin{tabular}{clrclr}
			\toprule
			\textbf{Variable Name} & \textbf{Range} & \textbf{Score} & \textbf{Variable Name} & \textbf{Range} & \textbf{Score} \\
			\midrule
			\multirow{6}[2]{*}{Age} & $<$39.5 & 0     & \multirow{6}[2]{*}{BMI} & $<$ 23.25 & 0 \\
			& 39.5$\thicksim$41.5 & 2.45  &       & 23.25$\thicksim$24.15 & 0.42 \\
			& 41.5$\thicksim$45.5 & 3.11  &       & 24.15$\thicksim$25.55 & 1.87 \\
			& 45.5$\thicksim$47.5 & 4.13  &       & 25.55$\thicksim$31.05 & 1.9 \\
			& 47.5$\thicksim$79.5 & 4.8   &       & 31.05$\thicksim$34.95 & 2.37 \\
			& $\ge$79.5 & 7.5   &       & $\ge$34.95 & 6.16 \\
			\midrule
			\multirow{5}[2]{*}{DBP} & $<$73   & 0     & \multirow{5}[2]{*}{TG} & $<$2.54 & 0 \\
			& 73$\thicksim$75 & 0.12  &       & 2.54$\thicksim$4.69 & 1.6 \\
			& 75$\thicksim$81 & 0.14  &       & 4.69$\thicksim$5.37 & 2.71 \\
			& 81$\thicksim$99 & 0.78  &       & 5.37$\thicksim$7.8 & 5.32 \\
			& $\ge$99  & 1.2   &       & $\ge$7.8 & 5.38 \\
			\midrule
			\multirow{4}[2]{*}{MCHC} & $<$337.5 & 0     & \multirow{4}[2]{*}{NEUT\%} & $<$56.85 & 0 \\
			& 337.5$\thicksim$342.5 & 0.14  &       & 56.85$\thicksim$69.35 & 0.19 \\
			& 342.5$\thicksim$351.5 & 0.65  &       & 69.35$\thicksim$77.65 & 0.89 \\
			& $\ge$351.5 & 1.32  &       & $\ge$77.65 & 1.68 \\
			\midrule
			\multirow{4}[2]{*}{PH} & $<$5.75 & 2.17  & \multirow{4}[2]{*}{PLT} & $<$131.5 & 1.08 \\
			& 5.75$\thicksim$5.25 & 2.1   &       & 131.5$\thicksim$205.5 & 0.75 \\
			& 5.25$\thicksim$6.25 & 1.13  &       & 205.5$\thicksim$342 & 0 \\
			& $\ge$6.25 & 0     &       & $\ge$342 & 3.01 \\
			\midrule
			\multirow{4}[2]{*}{BUN,UREAN} & $<$5.45 & 0     & \multirow{4}[2]{*}{LDL-C} & $<$1.5  & 0.33 \\
			& 5.45$\thicksim$5.85 & 0.24  &       & 1.5$\thicksim$3.44 & 0 \\
			& 5.85$\thicksim$6.55 & 0.84  &       & 3.44$\thicksim$4.64 & 0.75 \\
			& $\ge$6.55 & 1.37  &       & $\ge$4.64 & 2.26 \\
			\midrule
			\multirow{4}[2]{*}{Cr,CRE} & $<$43.5 & 4.36  & \multirow{4}[2]{*}{HDL-C} & $<$0.73 & 6.92 \\
			& 43.5$\thicksim$48.5 & 4.31  &       & 0.73$\thicksim$1.46 & 0.8 \\
			& 48.5$\thicksim$42.5 & 2.95  &       & 1.46$\thicksim$2.62 & 0 \\
			& $\ge$42.5 & 0     &       & $\ge$2.62 & 3.27 \\
			\midrule
			\multirow{6}[2]{*}{CEA} & $<$1.22 & 0     & \multirow{6}[2]{*}{WBC} & $<$3    & 4.85 \\
			& 1.22$\thicksim$2.18 & 1.73  &       & 3$\thicksim$6.29 & 0 \\
			& 2.18$\thicksim$5.05 & 2.45  &       & 6.29$\thicksim$7.24 & 0.24 \\
			& 5.05$\thicksim$6.95 & 3.86  &       & 7.24$\thicksim$7.7 & 0.24 \\
			& 6.95$\thicksim$9.21 & 6.98  &       & 7.7$\thicksim$10.19 & 0.65 \\
			& $\ge$9.21 & 7.6   &       & $\ge$10.19 & 1.67 \\
			\midrule
			\multirow{6}[2]{*}{AST/ALT} & $<$0.65 & 2.02  & \multirow{6}[2]{*}{UA} & $<$136  & 5.55 \\
			& 0.65$\thicksim$0.71 & 1.47  &       & 136$\thicksim$304.5 & 3.13 \\
			& 0.71$\thicksim$0.83 & 0.99  &       & 304.5$\thicksim$317.5 & 1.46 \\
			& 0.83$\thicksim$1.03 & 0.93  &       & 317.5$\thicksim$520.5 & 0 \\
			& \multirow{2}[1]{*}{$\ge$1.03} & \multirow{2}[1]{*}{0} &       & 520.5$\thicksim$539 & 1.54 \\
			&       &       &       & $\ge$539 & 1.72 \\
			\hline  
		\end{tabular}
	\end{threeparttable}
	}
	\label{score15}
\end{table}


\section{Discussion and Conclusion}
\label{last}
In this paper, we propose a fusion penalized logistic threshold regression model. The proposed model is flexible to account for nonlinear relationships between risk factors and the response using threshold regression framework. The fusion penalization encourages the preservation of some continuity of the risk score with respect to the levels of risk factors and reduce the total variance. A CART-type estimator is proposed to obtain the unknown threshold points, for which both consistency and non-asymptotic convergence rates are established. Using the non-asymptotic results for estimated threshold points, we deliver the consistency and selection guarantees of regression coefficients. Extensive simulation studies have shown the satisfactory performance of the proposed method. And we find it quite suitable for developing risk scores in diabetes prediction. However, inference on the threshold points and determination of the levels are still largely missing in literature. On the other hand, exploration of different types of risk scores based on the FILTER is of statistical interest itself. We put these lines of researches in the future. All the proofs for theorems and extra simulation results can be found at the online supplementary materials.




\bibliography{mybibfile}

\begin{thebibliography}{43}
\providecommand{\natexlab}[1]{#1}
\providecommand{\url}[1]{\texttt{#1}}
\expandafter\ifx\csname urlstyle\endcsname\relax
  \providecommand{\doi}[1]{doi: #1}\else
  \providecommand{\doi}{doi: \begingroup \urlstyle{rm}\Url}\fi

\bibitem[Abachi et~al.(2018)Abachi, Hosseini, Maskouni, Kangavari, and
  Cheung]{Abachi:etal:2018}
H.~M. Abachi, S.~Hosseini, M.~A. Maskouni, M.~Kangavari, and N.-M. Cheung.
\newblock Statistical discretization of continuous attributes using
  kolmogorov-smirnov test.
\newblock In \emph{Australasian Database Conference}, pages 309--315. Springer,
  2018.

\bibitem[Alaya et~al.(2019)Alaya, Bussy, Ga{\"\i}ffas, and
  Guilloux]{Alaya:etal:2019}
M.~Z. Alaya, S.~Bussy, S.~Ga{\"\i}ffas, and A.~Guilloux.
\newblock Binarsity: a penalization for one-hot encoded features in linear
  supervised learning.
\newblock \emph{J. Mach. Learn. Res.}, 20\penalty0 (118):\penalty0 1--34, 2019.

\bibitem[Bach(2010)]{Bach:2010}
F.~Bach.
\newblock Self-concordant analysis for logistic regression.
\newblock \emph{Electronic Journal of Statistics}, 4:\penalty0 384--414, 2010.

\bibitem[Banerjee and McKeague(2007)]{Banerjee:McKeague:2007}
M.~Banerjee and I.~W. McKeague.
\newblock Confidence sets for split points in decision trees.
\newblock \emph{The Annals of Statistics}, 35\penalty0 (2):\penalty0 543--574,
  2007.

\bibitem[Breiman et~al.(1984)Breiman, Friedman, Stone, and
  Olshen]{Breiman:etal:1984}
L.~Breiman, J.~H. Friedman, C.~J. Stone, and R.~A. Olshen.
\newblock \emph{Classification and regression trees}.
\newblock CRC Press, Boca Raton, 1984.

\bibitem[B{\"u}hlmann and Van De~Geer(2011)]{Buhlmann:Sara:2011}
P.~B{\"u}hlmann and S.~Van De~Geer.
\newblock \emph{Statistics for high-dimensional data: methods, theory and
  applications}.
\newblock Springer Science \& Business Media, 2011.

\bibitem[Bunea(2008)]{bunea2008honest}
F.~Bunea.
\newblock Honest variable selection in linear and logistic regression models
  via $\ell_{1}$ and $\ell_{1}$+ $\ell_{2}$ penalization.
\newblock \emph{Electronic Journal of Statistics}, 2:\penalty0 1153--1194,
  2008.

\bibitem[Chan(1993)]{Chan:1993}
K.-S. Chan.
\newblock Consistency and limiting distribution of the least squares estimator
  of a threshold autoregressive model.
\newblock \emph{The annals of statistics}, pages 520--533, 1993.

\bibitem[Dagenais(1969)]{Dagenais:1969}
M.~G. Dagenais.
\newblock A threshold regression model.
\newblock \emph{Econometrica: Journal of Econometric Society}, pages 193--203,
  1969.

\bibitem[Dussaut et~al.(2017)Dussaut, Gallo, Carballido, and
  Ponzoni]{Dussaut:etal:2017}
J.~S. Dussaut, C.~A. Gallo, J.~A. Carballido, and I.~Ponzoni.
\newblock Analysis of gene expression discretization techniques in microarray
  biclustering.
\newblock In \emph{International Conference on Bioinformatics and Biomedical
  Engineering}, pages 257--266. Springer, 2017.

\bibitem[Ehm et~al.(2016)Ehm, Gneiting, Jordan, and Kr{\"u}ger]{Ehm2016}
W.~Ehm, T.~Gneiting, A.~Jordan, and F.~Kr{\"u}ger.
\newblock Of quantiles and expectiles: consistent scoring functions, choquet
  representations and forecast rankings.
\newblock \emph{Journal of the Royal Statistical Society: Series B (Statistical
  Methodology)}, 78\penalty0 (3):\penalty0 505--562, 2016.

\bibitem[Ferreira and Figueiredo(2015)]{Ferreira:Figueiredo:2015}
A.~J. Ferreira and M.~A. Figueiredo.
\newblock Feature discretization with relevance and mutual information
  criteria.
\newblock In \emph{Pattern recognition applications and methods}, pages
  101--118. Springer, 2015.

\bibitem[Fithian and Hastie(2014)]{FH14}
W.~Fithian and T.~Hastie.
\newblock Local case-control sampling: Efficient subsampling in imbalanced data
  sets.
\newblock \emph{Annals of statistics}, 42\penalty0 (5):\penalty0 1693--1724,
  2014.

\bibitem[Flores et~al.(2019)Flores, Calvo, and Perez]{Flores2019}
J.~L. Flores, B.~Calvo, and A.~Perez.
\newblock Supervised non-parametric discretization based on kernel density
  estimation.
\newblock \emph{Pattern Recognition Letters}, 128:\penalty0 496--504, 2019.

\bibitem[Franc et~al.(2018)Franc, Fikar, Bartos, and Sofka]{Franc:etal:2017}
V.~Franc, O.~Fikar, K.~Bartos, and M.~Sofka.
\newblock Learning data discretization via convex optimization.
\newblock \emph{Machine Learning}, 107\penalty0 (2):\penalty0 333--355, 2018.

\bibitem[Friedman(1991)]{Friedman:1991}
J.~H. Friedman.
\newblock Multivariate adaptive regression splines.
\newblock \emph{The annals of statistics}, 19\penalty0 (1):\penalty0 1--67,
  1991.

\bibitem[Fu et~al.(2017)Fu, Liu, Jiang, Wu, and Hsu]{Fu:etal:2017}
B.~Fu, H.~Liu, Z.~Jiang, Z.~Wu, and D.~F. Hsu.
\newblock D-fs: A novel integration method of discretization and feature
  selection.
\newblock In \emph{2017 14th International Symposium on Pervasive Systems,
  Algorithms and Networks \& 2017 11th International Conference on Frontier of
  Computer Science and Technology \& 2017 Third International Symposium of
  Creative Computing (ISPAN-FCST-ISCC)}, pages 6--13. IEEE, 2017.

\bibitem[Gao et~al.(2013)Gao, Tj{\o}stheim, and Yin]{Gao:etal:2013}
J.~Gao, D.~Tj{\o}stheim, and J.~Yin.
\newblock Estimation in threshold autoregressive models with a stationary and a
  unit root regime.
\newblock \emph{Journal of Econometrics}, 172\penalty0 (1):\penalty0 1--13,
  2013.

\bibitem[Garcia et~al.(2012)Garcia, Luengo, S{\'a}ez, Lopez, and
  Herrera]{Garcia:etal:2013}
S.~Garcia, J.~Luengo, J.~A. S{\'a}ez, V.~Lopez, and F.~Herrera.
\newblock A survey of discretization techniques: Taxonomy and empirical
  analysis in supervised learning.
\newblock \emph{IEEE transactions on Knowledge and Data Engineering},
  25\penalty0 (4):\penalty0 734--750, 2012.

\bibitem[Gneiting and Raftery(2007)]{GR:2007}
T.~Gneiting and A.~E. Raftery.
\newblock Strictly proper scoring rules, prediction, and estimation.
\newblock \emph{Journal of the American statistical Association}, 102\penalty0
  (477):\penalty0 359--378, 2007.

\bibitem[Hansen(2000)]{Hansen:2000}
B.~E. Hansen.
\newblock Sample splitting and threshold estimation.
\newblock \emph{Econometrica}, 68\penalty0 (3):\penalty0 575--603, 2000.

\bibitem[Ishwaran(2015)]{Ishwaran:2015}
H.~Ishwaran.
\newblock The effect of splitting on random forests.
\newblock \emph{Machine learning}, 99\penalty0 (1):\penalty0 75--118, 2015.

\bibitem[Luo et~al.(2014)Luo, Han, Zeng, Chen, Pan, Wang, and Zhang]{Plos}
S.~Luo, L.~Han, P.~Zeng, F.~Chen, L.~Pan, S.~Wang, and T.~Zhang.
\newblock A risk assessment model for type 2 diabetes in chinese.
\newblock \emph{PloS one}, 9\penalty0 (8):\penalty0 e104046, 2014.

\bibitem[M{\'a}rquez-Grajales et~al.(2020)M{\'a}rquez-Grajales, Acosta-Mesa,
  Mezura-Montes, and Graff]{Marquez:etal:2020}
A.~M{\'a}rquez-Grajales, H.-G. Acosta-Mesa, E.~Mezura-Montes, and M.~Graff.
\newblock A multi-breakpoints approach for symbolic discretization of time
  series.
\newblock \emph{Knowledge and Information Systems}, 62\penalty0 (7):\penalty0
  2795--2834, 2020.

\bibitem[McClish(1989)]{McClish:1989}
D.~K. McClish.
\newblock Analyzing a portion of the roc curve.
\newblock \emph{Medical Decision Making}, 9\penalty0 (3):\penalty0 190--195,
  1989.
\newblock \doi{10.1177/0272989X8900900307}.

\bibitem[McCullagh and Nelder(1989)]{McCullagh:Nelder:1989}
P.~McCullagh and J.~A. Nelder.
\newblock \emph{Generalized linear models}.
\newblock Chapman and Hall, 1989.

\bibitem[Noble et~al.(2011)Noble, Mathur, Dent, Meads, and
  Greenhalgh]{Nobel2011}
D.~Noble, R.~Mathur, T.~Dent, C.~Meads, and T.~Greenhalgh.
\newblock Risk models and scores for type 2 diabetes: systematic review.
\newblock \emph{Bmj}, 343, 2011.

\bibitem[Park and Hastie(2007)]{Park:Hastie:2007}
M.~Y. Park and T.~Hastie.
\newblock L1-regularization path algorithm for generalized linear models.
\newblock \emph{Journal of the Royal Statistical Society: Series B (Statistical
  Methodology)}, 69\penalty0 (4):\penalty0 659--677, 2007.

\bibitem[Petersen et~al.(2016)Petersen, Witten, and Simon]{Petersen:etal:2016}
A.~Petersen, D.~Witten, and N.~Simon.
\newblock Fused lasso additive model.
\newblock \emph{Journal of Computational and Graphical Statistics}, 25\penalty0
  (4):\penalty0 1005--1025, 2016.

\bibitem[Prentice and Pyke(1979)]{PP79}
R.~L. Prentice and R.~Pyke.
\newblock Logistic disease incidence models and case-control studies.
\newblock \emph{Biometrika}, 66\penalty0 (3):\penalty0 403--411, 1979.

\bibitem[Qian(1998)]{Qian:1998}
L.~Qian.
\newblock On maximum likelihood estimators for a threshold autoregression.
\newblock \emph{Journal of Statistical Planning and Inference}, 75\penalty0
  (1):\penalty0 21--46, 1998.

\bibitem[Ravikumar et~al.(2010)Ravikumar, Wainwright, and
  Lafferty]{Ravikumar:etal:2010}
P.~Ravikumar, M.~J. Wainwright, and J.~D. Lafferty.
\newblock High-dimensional ising model selection using $\ell_{1}$-regularized
  logistic regression.
\newblock \emph{The Annals of Statistics}, 38\penalty0 (3):\penalty0
  1287--1319, 2010.

\bibitem[Robin et~al.(2011)Robin, Turck, Hainard, Tiberti, Lisacek, Sanchez,
  and M{\"u}ller]{Robin:etal:2011}
X.~Robin, N.~Turck, A.~Hainard, N.~Tiberti, F.~Lisacek, J.-C. Sanchez, and
  M.~M{\"u}ller.
\newblock proc: an open-source package for r and s+ to analyze and compare roc
  curves.
\newblock \emph{BMC bioinformatics}, 12\penalty0 (1):\penalty0 1--8, 2011.

\bibitem[Sokolovska et~al.(2018)Sokolovska, Chevaleyre, and
  Zucker]{Sokolovska:etal:2018}
N.~Sokolovska, Y.~Chevaleyre, and J.-D. Zucker.
\newblock A provable algorithm for learning interpretable scoring systems.
\newblock In \emph{International Conference on Artificial Intelligence and
  Statistics}, pages 566--574. PMLR, 2018.

\bibitem[Sriwanna et~al.(2019)Sriwanna, Boongoen, and
  Iam-On]{Sriwanna:etal:2019}
K.~Sriwanna, T.~Boongoen, and N.~Iam-On.
\newblock Graph clustering-based discretization approach to microarray data.
\newblock \emph{Knowledge and Information Systems}, 60\penalty0 (2):\penalty0
  879--906, 2019.

\bibitem[Sur and Cand{\`e}s(2019)]{sur2019modern}
P.~Sur and E.~J. Cand{\`e}s.
\newblock A modern maximum-likelihood theory for high-dimensional logistic
  regression.
\newblock \emph{Proceedings of the National Academy of Sciences}, 116\penalty0
  (29):\penalty0 14516--14525, 2019.

\bibitem[Tang and Song(2016)]{Tang:Song:2016}
L.~Tang and P.~X. Song.
\newblock Fused lasso approach in regression coefficients clustering: learning
  parameter heterogeneity in data integration.
\newblock \emph{The Journal of Machine Learning Research}, 17\penalty0
  (1):\penalty0 3915--3937, 2016.

\bibitem[Tibshirani et~al.(2005)Tibshirani, Saunders, Rosset, Zhu, and
  Knight]{Tibshirani:etal:2005}
R.~Tibshirani, M.~Saunders, S.~Rosset, J.~Zhu, and K.~Knight.
\newblock Sparsity and smoothness via the fused lasso.
\newblock \emph{Journal of the Royal Statistical Society: Series B (Statistical
  Methodology)}, 67\penalty0 (1):\penalty0 91--108, 2005.

\bibitem[Tsai and Chen(2019)]{Tsai:Chen:2019}
C.-F. Tsai and Y.-C. Chen.
\newblock The optimal combination of feature selection and data discretization:
  An empirical study.
\newblock \emph{Information Sciences}, 505:\penalty0 282--293, 2019.

\bibitem[Vollmer et~al.(2019)Vollmer, Golab, B{\"o}hm, and
  Srivastava]{Vollmer:etal:2019}
M.~Vollmer, L.~Golab, K.~B{\"o}hm, and D.~Srivastava.
\newblock Informative summarization of numeric data.
\newblock In \emph{Proceedings of the 31st International Conference on
  Scientific and Statistical Database Management}, pages 97--108, 2019.

\bibitem[Wainwright(2009)]{Wainwright:2009}
M.~J. Wainwright.
\newblock Sharp thresholds for high-dimensional and noisy sparsity recovery
  using $\ell_{1}$-constrained quadratic programming (lasso).
\newblock \emph{IEEE transactions on information theory}, 55\penalty0
  (5):\penalty0 2183--2202, 2009.

\bibitem[Wang(2016)]{wang2018}
L.~Wang.
\newblock \emph{Some topics on model-based clustering}.
\newblock PhD thesis, Colorado State University, 2016.

\bibitem[Zhao and Yu(2006)]{Zhao:Yu:2004}
P.~Zhao and B.~Yu.
\newblock On model selection consistency of lasso.
\newblock \emph{The Journal of Machine Learning Research}, 7:\penalty0
  2541--2563, 2006.

\end{thebibliography}



\end{document}